\documentclass[sigconf]{acmart}

\usepackage{url}
\usepackage{threeparttable}
\usepackage{multirow} 
\usepackage{makecell}
\usepackage{algorithm, algorithmic}
\usepackage{setspace}
\usepackage{subcaption}
\usepackage{listings}
\usepackage{xcolor}
\usepackage{pgffor}
\usepackage{wrapfig}
\usepackage{tikzsymbols}
\usepackage{wrapfig}
\usepackage{mathtools}
\usepackage{bm}
\usepackage{pifont}

\AtBeginDocument{%
  }

\setcopyright{acmlicensed}
\copyrightyear{2026}
\acmYear{2026}
\acmDOI{XXXXXXX.XXXXXXX}
\acmConference[KDD '26]{Proceedings of the 28th ACM SIGKDD Conference on Knowledge Discovery and Data Mining}{August 09--13,
  2026}{Jeju, Korea}
\acmISBN{978-1-4503-XXXX-X/2026/08}

\acmSubmissionID{1186}

\theoremstyle{definition}
\newtheorem{theorem}{Theorem}
\newtheorem{remark}{Remark}
\newtheorem{definition}{Definition}

\begin{document}

\def\ie{\textit{i.e.}}
\def\eg{\textit{e.g.}}

\title{Rank Matters: Understanding and Defending Model Inversion Attacks via Low-Rank Feature Filtering}


\author{Hongyao Yu}
\authornote{Authors contributed equally to this research.}
\orcid{0009-0009-8525-1565}
\affiliation{%
  \institution{Tsinghua Shenzhen International Graduate School, Tsinghua University}
  \city{Shenzhen}
  \state{Guangdong}
  \country{China}
}
\email{chrisqcwx@gmail.com}

\author{Yixiang Qiu}
\authornotemark[1]
\orcid{0009-0000-3444-5807}
\affiliation{%
  \institution{Tsinghua Shenzhen International Graduate School, Tsinghua University}
  \city{Shenzhen}
  \state{Guangdong}
  \country{China}
}
\email{qiu-yx24@mails.tsinghua.edu.cn}

\author{Hao Fang}
\authornotemark[1]
\orcid{0009-0004-0271-6579}
\affiliation{%
  \institution{Tsinghua Shenzhen International Graduate School, Tsinghua University}
  \city{Shenzhen}
  \state{Guangdong}
  \country{China}
}
\email{fang-h23@mails.tsinghua.edu.cn}

\author{Tianqu Zhuang}
\authornotemark[1]
\orcid{0009-0006-8730-1878}
\affiliation{%
  \institution{Tsinghua Shenzhen International Graduate School, Tsinghua University}
  \city{Shenzhen}
  \state{Guangdong}
  \country{China}
}
\email{zhuangtq23@mails.tsinghua.edu.cn}

\author{Bin Chen}
\authornote{Corresponding author.}
\orcid{0000-0002-4798-230X}
\affiliation{%
  \institution{Harbin Institute of Technology, Shenzhen}
  \city{Shenzhen}
  \state{Guangdong}
  \country{China}
}
\email{chenbin2021@hit.edu.cn}

\author{Sijin Yu}
\orcid{0009-0005-2352-4733}
\affiliation{%
  \institution{South China University of Technology}
  \city{Guangzhou}
  \state{Guangdong}
  \country{China}
}
\email{eeyusijin@mail.scut.edu.cn}

\author{Bin Wang}
\orcid{0009-0004-9725-5305}
\affiliation{%
  \institution{Tsinghua Shenzhen International Graduate School, Tsinghua University}
  \city{Shenzhen}
  \state{Guangdong}
  \country{China}
}
\email{18922260521@189.cn}

\author{Shu-Tao Xia}
\orcid{0000-0002-8639-982X}
\affiliation{%
  \institution{Tsinghua Shenzhen International Graduate School, Tsinghua University}
  \city{Shenzhen}
  \state{Guangdong}
  \country{China}
}
\email{xiast@sz.tsinghua.edu.cn}

\author{Ke Xu}
\orcid{0000-0003-2587-8517}
\affiliation{%
  \institution{Department of Computer Science and Technology, Tsinghua University}
  \state{Beijing}
  \country{China}
}
\email{xuke@tsinghua.edu.cn}

\renewcommand{\shortauthors}{Hongyao Yu et al.}

\begin{abstract}

Model Inversion Attacks (MIAs) pose a significant threat to data privacy by reconstructing sensitive training samples from the knowledge embedded in trained machine learning models. Despite recent progress in enhancing the effectiveness of MIAs across diverse settings, defense strategies have lagged behind—struggling to balance model utility with robustness against increasingly sophisticated attacks. In this work, we propose the ideal inversion error to measure the privacy leakage, and our theoretical and empirical investigations reveals that higher-rank features are inherently more prone to privacy leakage. Motivated by this insight, we propose a lightweight and effective defense strategy based on low-rank feature filtering, which explicitly reduces the attack surface by constraining the dimension of intermediate representations. Extensive experiments across various model architectures and datasets demonstrate that our method consistently outperforms existing defenses, achieving state-of-the-art performance against a wide range of MIAs. Notably, our approach remains effective even in challenging regimes involving high-resolution data and high-capacity models, where prior defenses fail to provide adequate protection. The code is available at \url{https://github.com/Chrisqcwx/LoFt}.

\end{abstract}




\maketitle

\section{Introduction}

{

\begin{figure*}[!htbp]
    \begin{subfigure}[b]{\textwidth}
    \centering
    \begin{minipage}[t]{0.33\textwidth}
                \centering
        \includegraphics[width=\textwidth]{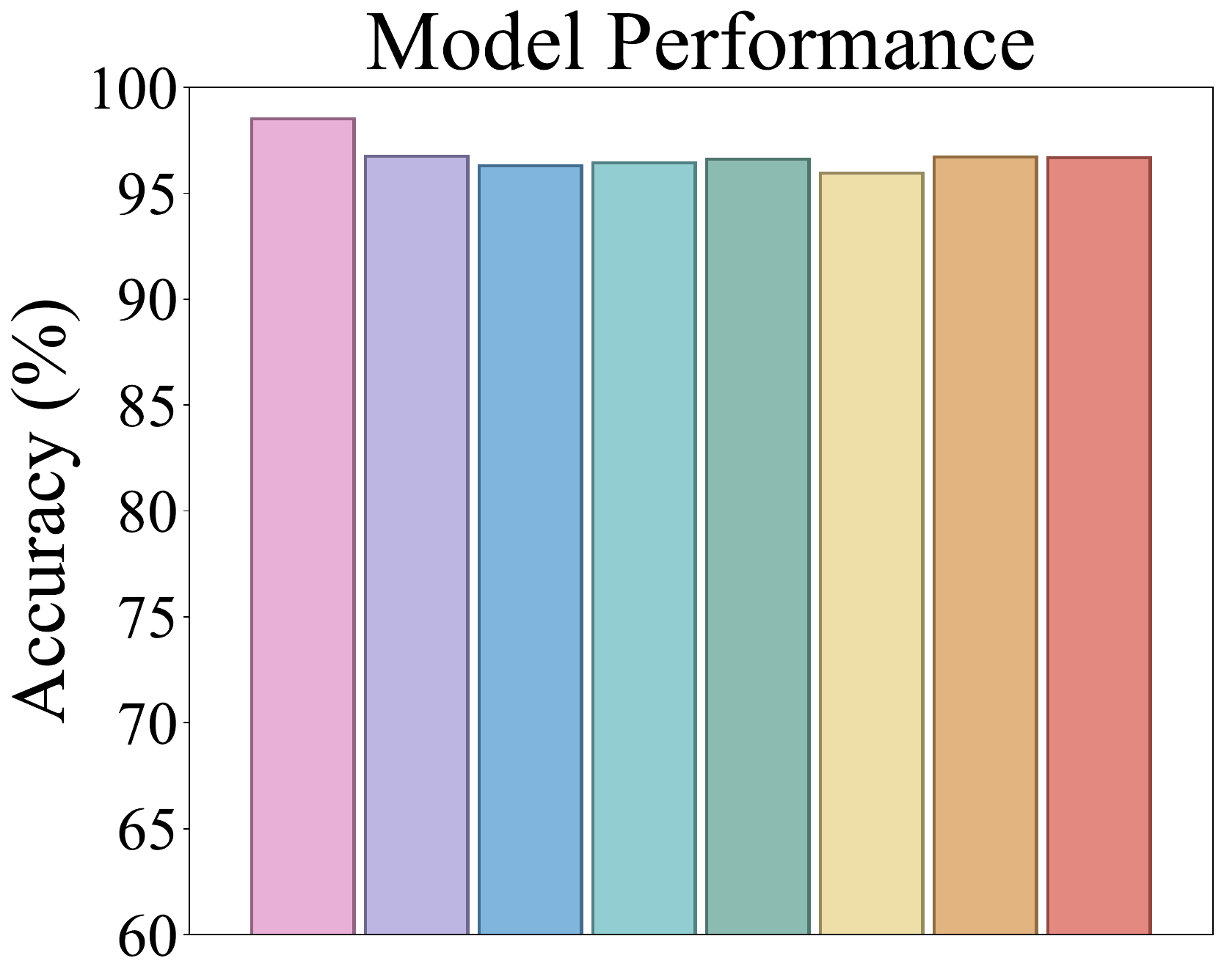}
    \end{minipage}%
    \begin{minipage}[t]{0.33\textwidth}
                \centering
        \includegraphics[width=\textwidth]{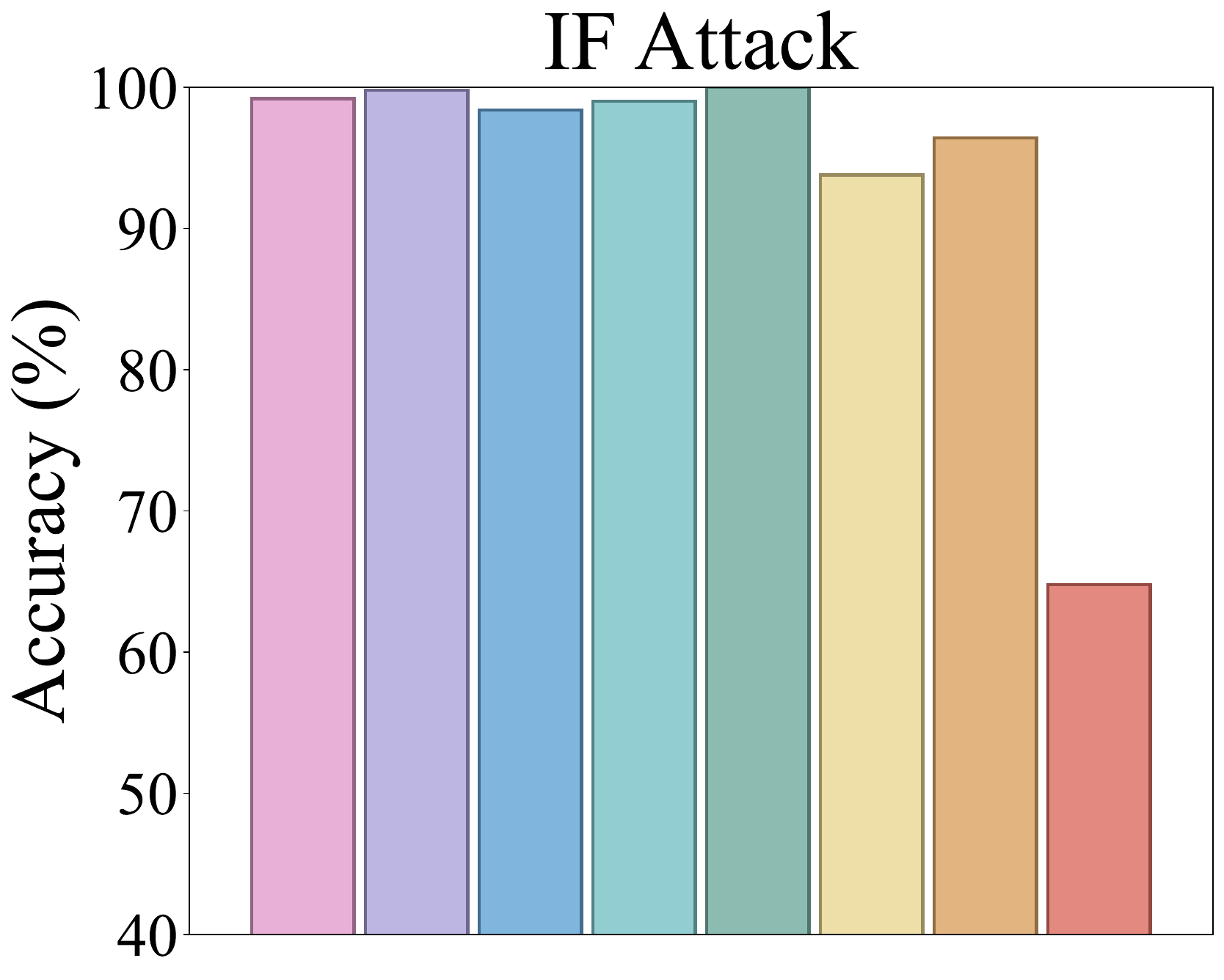}
    \end{minipage}%
    \begin{minipage}[t]{0.33\textwidth}
                \centering
        \includegraphics[width=\textwidth]{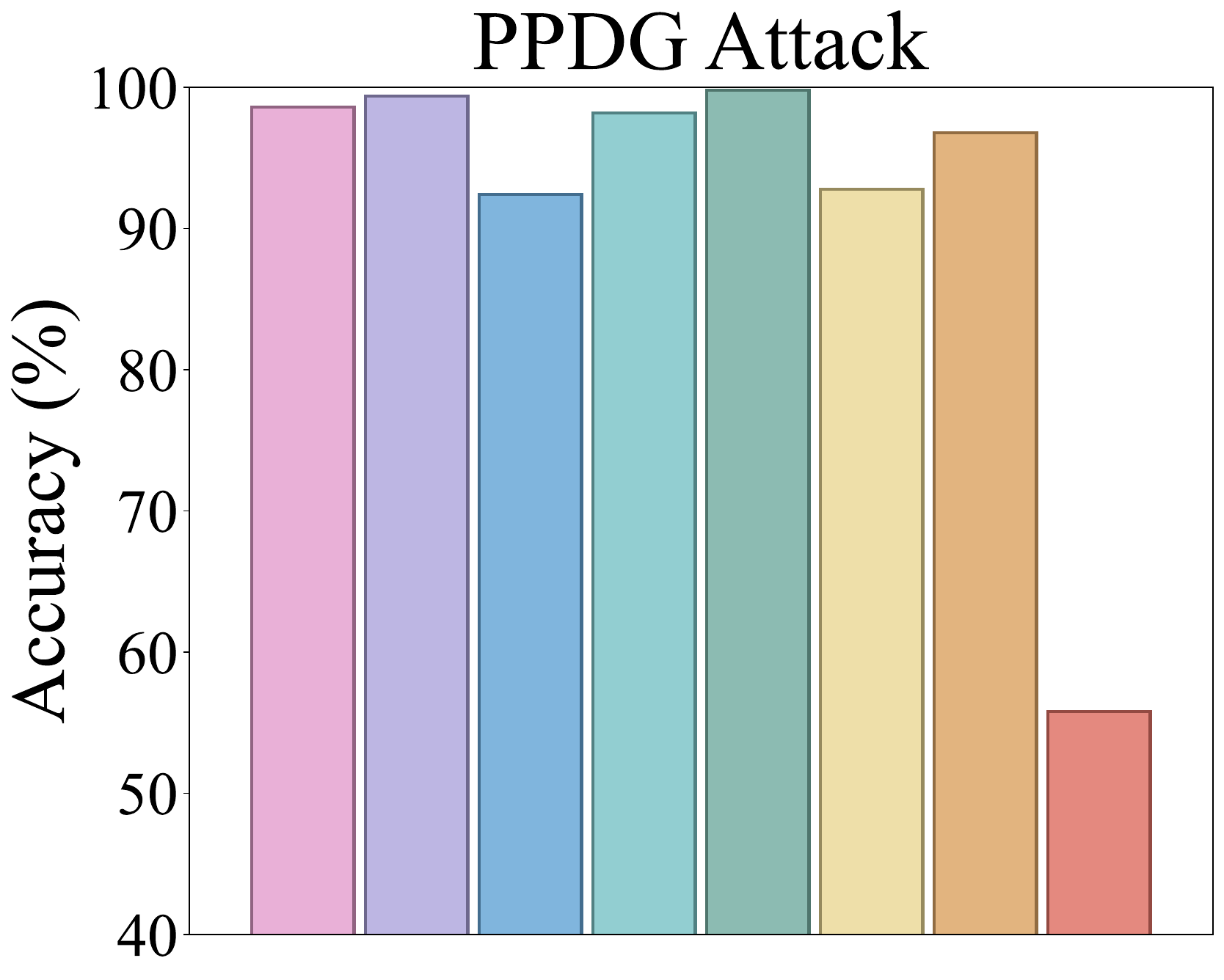}
    \end{minipage}%
    \end{subfigure}
    \centering
    \begin{subfigure}[b]{\textwidth}
    \begin{minipage}[t]{\textwidth}
        \centering
        \includegraphics[width=\textwidth]{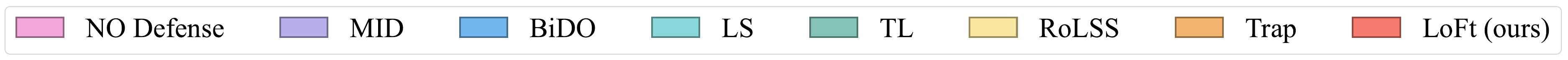}
    \end{minipage}
    \end{subfigure}
    \caption{Model performance on the test dataset and attack accuracy with difference defenses in the high-resolution scenarios. }
    \label{fig: intro high}
\end{figure*}

}

In recent years, Deep Neural Networks (DNNs) have made remarkable progress, achieving impressive performance across a wide range of applications, such as face recognition \cite{resnet,facenet112}, audio recognition \cite{audio}, and medical research \cite{medicine}.
However, the powerful capabilities that make these models so effective also render them vulnerable to privacy attacks \cite{fang2023gifd,qiurevisit,icas}.
One of the most critical threats to privacy and security is the Model Inversion Attack (MIA) \cite{mibench}, which allows adversaries to reconstruct privacy-sensitive training data from the output information of released models. For example, in face recognition systems, MIAs can produce synthetic images that reveal specific visual characteristics of private identities even without direct access to the private training dataset. This poses a significant risk, as MIAs may allow unauthorized individuals to reconstruct valid facial features to disguise themselves as authorized personnel, thereby compromising both privacy and security.

While MIAs have rapidly advanced in sophistication and effectiveness, existing defense methods lag behind, particularly against powerful attacks like IF \cite{ifgmi} and PPDG \cite{ppdg}. Moreover, recent defense researches \cite{mid,bido,trap} execute in the scenarios where the target model has relatively poor performance. However, as shown in Fig. \ref{fig: intro high}, when the target models achieve high predictive accuracy, \ie, more than $96\%$ accuracy on the test split of the FaceScrub dataset \cite{facescrub} in high-resolution scenarios, \textit{existing defenses are almost completely incapable of defending against advanced attacks.}


We start by proposing an Ideal Inversion Error (IIE) metric to quantify the risk of privacy leakage. We then theoretically analyze the relationship between IIE and the rank of weight matrix in the linear head, which indicates that a higher rank leads to more serious privacy leakage.

Motivated by this insight, to protect the private training data of classifier, we propose a novel \textbf{lo}w-rank \textbf{f}ea\textbf{t}ure filtering strategy (LoFt). The core 
idea is to reduce the rank of the linear classification head by decomposing it into two linear layers. The first layer compress the input features into a low dimension, and the second use the compressed feature to generate the output. This approach ensures a low rank in the classification head. This encourages the model to focus on task-relevant information while filtering out the redundant features, which is meaningless for the main task but essential for the reconstruction process of attackers. Additionally, we introduce a $\mathrm{Tanh}$ activation to induce gradient vanishing, further impeding the optimization process of inversion attacks.

We conduct comprehensive experiments across multiple settings to evaluate our method. The results demonstrate that our approach achieves state-of-the-art (SOTA) defense performance.
In scenarios where target models demonstrate strong performance, existing defense methods fail to resist advanced attacks, while our method maintains superior defense capabilities, as shown in Fig.~\ref{fig: intro high}.
Extensive experiments and ablation studies thoroughly validate the effectiveness of our proposed method. 

In summary, our main contributions are as follows:
\begin{itemize}
    \item We propose Ideal Inversion Error (IIE), a novel metric to quantify the risk of privacy leakage for any model. We theoretically analyze the relationship between IIE and the rank of weight matrix, for both linear models and deep neural networks.
    \item Based on our theory, we propose LoFt, a low-rank feature filtering defense against model inversion attacks. LoFt reduces the rank of the last weight matrix, thereby increasing the IIE of the model to enhance its robustness.
    \item We conduct extensive experiments to demonstrate that LoFt offers superior robustness against strong MIAs across multiple settings, particularly in high-performance and high-resolution regimes where prior defenses break down.
\end{itemize}

\section{Background and Related Works}

\subsection{Model Inversion Attacks}
\label{sec:bg/mia}

Let $f_\theta: \mathbb{X}\rightarrow [0, 1]^n$ denote a general classifier, which processes a private image $\bm{x}\sim\mathcal{P}(X)$ and computes a prediction $\hat{y}_c\in[0,1]$ for each class $c\in \{1, \dots, n\}$. 
MIAs aim to reconstruct images that reveal private characteristic features of a specific identity $c$.
In white-box scenarios, attackers have full access to the classifier's weights and outputs, allowing them to compute gradients across the classifier when executing the attack.

In recent years, Generative Adversarial Network (GAN) based model inversion attacks, first proposed by GMI \cite{gmi}, have become the standard paradigm for MIAs \cite{fang2024privacy}.
Specifically, attackers initially train a generator $G$ to capture a similar structural prior as the target private data. 
In the attack stage, they attempt to reconstruct private images $\bm{x}^*=G(\bm{z}^*)$ labeled with class $c$ to approximate the private distribution $\mathcal{P}(X)$, by optimizing the latent vectors $\bm{z}$:
\begin{equation}
    \bm{z}^*=\mathop{\arg\min}_{\bm{z}} \mathcal{L}_{cls}  (f_\theta(G(\bm{z})), c) + \lambda \mathcal{L}_{prior}(\bm{z}; G),
\end{equation}
where $\mathcal{L}_{cls}$ is the classification loss, $\lambda$ is a hyperparameter, and $\mathcal{L}_{prior}$ denotes the optional prior knowledge regularization terms, such as the discriminator loss \cite{gmi, ked} and the feature regularization loss \cite{lomma}. 

Subsequent MIA studies have largely followed this pipeline and improved upon it. 
%
KED \cite{ked} and PLG \cite{plg} propose to fully extract the private information learned by the target classifier. They leverage the target classifier to annotate pseudo-labels for unlabeled images to train target-specific GANs.
Additionally, some researchers have explored various classification losses to mitigate the effects of gradient vanishing issues, such as the Poincar\'e loss \cite{ppa,ifgmi,ppdg}, the max-margin loss \cite{plg} and the logit loss \cite{lomma}.
%
To enhance the attack performance in high-resolution scenarios, Mirror \cite{mirror}, PPA \cite{ppa}, IF \cite{ifgmi}, and PPDG \cite{ppdg} utilize well-structured StyleGANs \cite{stylegan2} to generate high-quality and high-resolution samples. Specifically, IF \cite{ifgmi} and PPDG 
\cite{ppdg} achieve state-of-the-art (SOTA) attack performance, posing a serious threat to privacy.

\subsection{Model Inversion Defenses}
\label{sec:bg/midefense}
Model inversion defenses aim at reducing the threat of MIAs.
\cite{mid} proposes the first MI-specific defense method MID, defending attacks by reducing the mutual information between model input and output. 
\cite{bido} upgrades from MID with a bilateral dependency optimization (BiDO). They minimize the dependency between model inputs and the intermediate representations while maximizing that between the intermediate representations and model outputs. 
\cite{ls} explores the connection between the label smoothing technique and model robustness against MIAs. They indicate that the cross-entropy loss with a negative label smoothing factor can enhance the model robustness and vice versa. 
\cite{tl} finds that the parameters in the first few layers are essential for model inversion attacks and the last ones are most critical for classification tasks. Therefore, they first pre-trains the model on public datasets. Then, they freeze the parameters in previous layers when fine-tuning on private datasets.
\cite{rolss} proposes to modify the model architecture to introduce gradient vanishing for attackers. They remove the skip connections in the last few residual layers. 
\cite{trap} actively injects a backdoor into the model, guiding the attacker to reconstruct the backdoor features instead of the privacy features.
\cite{zhuang2025stealthy} proposes a defense against black-box MIAs based on conditional mutual information.
More details about the defenses are described in Appendix \ref{appx:defense summary}.
However, recent defense methods are no longer able to resist the most advanced MIAs, such as PPDG \cite{ppdg} and IF \cite{ifgmi}.

\section{A Larger Rank Leads to More Privacy Leakage}
\label{sec:lowrank motivation}


Modern machine learning models, particularly deep neural networks, often rely on a large number of intermediate features and high-rank dense layers to enhance expressiveness and performance. However, this increased representational capacity can also make it easier for attackers to infer sensitive information, thereby increasing the risk of privacy leakage.



To rigorously characterize the privacy risks associated with model inversion attacks, we introduce the concept of the inversion problem. Let $f_\theta: \mathbb{X}\rightarrow \mathbb{Y}$ be a model with input $\bm{x}\sim \mathcal{P}_{\bm{x}}$ and define the model output as $\bm{y}=f_\theta (\bm{x})$. Inversion attacks aim to reconstruct the input $\bm{x}$ based on knowledge of the model $f_\theta$ and its output $\bm{y}$.



We consider the strongest possible adversary—one with full knowledge of the model and the data distribution. This motivates the definition of the Ideal Inversion Attacker (IIA), which reconstructs the input by minimizing the expected reconstruction error conditioned on the observed output.

\begin{definition}[Ideal Inversion Attacker, IIA]
Let $e\colon\mathbb{X}\times\mathbb{X}\to\mathbb{R}_{\geq0}$ be an error function. For any $\bm{y}\in f_\theta (\bm{\mathbb{X}})$, an IIA always yields
\begin{align}
\label{eq:理想攻击者}
\mathop{\arg\min}_{\hat{\bm{x}}\in\mathbb{X}}\mathop{\mathbb{E}}_{\bm{x}\sim\mathcal{P}_{\bm{x}}}\big[e(\hat{\bm{x}},\bm{x})\big| f_\theta (\bm{x})=\bm{y}\big]
\end{align}
which is the estimation of $\bm{x}$ that minimizes the expected error under the posterior distribution given the model output.
\end{definition}

\begin{remark}
In practice, attackers typically lack access to the true data distribution $\mathcal{P}_{\bm{x}}$. Hence, the IIA serves as a theoretical construct that provides a principled lower bound on the achievable reconstruction error.
\end{remark}

To quantify the privacy risk of a model, we define the Ideal Inversion Error (IIE), which measures the expected reconstruction error incurred by the IIA over the input distribution.

\begin{definition}[Ideal Inversion Error, IIE]
\begin{align}
\text{IIE}(f_\theta)=\mathop{\mathbb{E}}_{\bm{y}\sim\mathcal{P}_{f_\theta (\bm{x})}}\Big[\mathop{\min}_{\hat{\bm{x}}\in\mathbb{X}}\mathop{\mathbb{E}}_{\bm{x}\sim\mathcal{P}_{\bm{x}}}\big[e(\hat{\bm{x}},\bm{x})\big| f_\theta (\bm{x})= \bm{y}\big]\Big],
\end{align}
which is equivalent to 
\begin{align}
\text{IIE}(f_\theta)=\mathop{\mathbb{E}}_{\bm{x'}\sim\mathcal{P}_{\bm{x}}}\Big[\mathop{\min}_{\hat{\bm{x}}\in\mathbb{X}}\mathop{\mathbb{E}}_{\bm{x}\sim\mathcal{P}_{\bm{x}}}\big[e(\hat{\bm{x}},\bm{x})\big| f_\theta (\bm{x})= f_\theta (\bm{x'})\big]\Big].
\end{align}
\end{definition}

\begin{remark}
The IIE quantifies the expected uncertainty an IIA faces when reconstructing inputs. It thus provides a principled metric for evaluating the privacy robustness of a model $ f_\theta $.
\end{remark}



We now focus on a specific instantiation of the model, a linear mapping. Let the model be $ f_\theta (\bm{x})=\bm{W}\bm{x}$ where $\bm{W} \in \mathbb{R}^{n \times m}$. This setting is motivated by two basic facts:
\begin{enumerate}
    \item A linear layer is basically the last layer of the classifier \cite{resnet,facenet112}.
    \item The input of this last layer, which correspond to feature embeddings of private samples, encodes sensitive sematic information of the private samples, Thus it requires protection \cite{gmi,ppa}.
\end{enumerate}
For simplicity, we initially set the input distribution of this linear model to be a standard normal distribution, \ie, $\mathcal{P}_{\bm{x}}= \mathcal{N}(\bm{0}_m,\bm{I}_{m\times m})$,
and we set $e(\hat{\bm{x}},\bm{x})=||\hat{\bm{x}}-\bm{x}||_2^2$.

\begin{theorem}
Under the above assumptions, the Ideal Inversion Error of the linear model $f_\theta(\bm{x})=\bm{Wx}$ is given by
\begin{align}
\text{IIE}(f_\theta)=m-\text{rank}(\bm{W}).
\end{align}
\end{theorem}

\begin{proof}
With the assumption of the error function and input distribution, the IIE can be calculated as:
\begin{align}
\text{IIE}(f_\theta)
=&\mathop{\mathbb{E}}_{\bm{x}'\sim\mathcal{P}_{\bm{x}}}\Big[\mathop{\min}_{\hat{\bm{x}}\in\mathbb{X}}\mathop{\mathbb{E}}_{\bm{x}\sim\mathcal{P}_{\bm{x}}}\big[\Vert\hat{\bm{x}}-\bm{x}\Vert^2_2\big|\bm{W}\bm{x}=\bm{W}\bm{x}'\big]\Big],\\
=&\mathop{\mathbb{E}}_{\bm{x}'\sim\mathcal{P}_{\bm{x}}}\Big[\mathop{\min}_{\hat{\bm{x}}\in\mathbb{X}}\mathop{\mathbb{E}}_{\bm{x}\sim\mathcal{P}_{\bm{x}}}\big[\Vert\hat{\bm{x}}-\bm{x}\Vert^2_2\big|\bm{x}\in\text{null}(\bm{W})+\bm{x}'\big]\Big],\\
=&\mathop{\mathbb{E}}_{\bm{x}'\sim\mathcal{P}_{\bm{x}}}\Big[\mathop{\mathrm{Var}}_{\bm{x}\sim\mathcal{P}_{\bm{x}}}\big[\bm{x}\big|\bm{x}\in\text{null}(\bm{W})+\bm{x}'\big]\Big].
\end{align}
%
Let $S=\text{null}(\bm{W})+\bm{x}'\subseteq\mathbb{R}^m$ denote the affine subspace consistent with the observed output. Since $\bm{x}\sim \mathcal{N}(\bm{0}_m,\bm{I}_{m\times m})$, the conditional distribution of $\bm{x}\in S$ is standard Gaussian restricted to the affine subspace $S$.  Therefore, we have
\begin{align}
\text{IIE}(f_\theta)
&=\mathop{\mathbb{E}}_{\bm{x}'\sim\mathcal{P}_{\bm{x}}}\big[\text{dim}(S)\big]=\text{dim}(S)=m-\text{rank}(\bm{W}).
\end{align}
\end{proof}




When considering a deep neural network (DNN) of the form $f_\theta=\mathcal{C}\circ \mathcal{E}$, we decompose it into two components without loss of generality: a non-linear encoder \( \mathcal{E} \) maps input images to feature representations, and a linear head \( \mathcal{C} \). 
Empirical evidence from \cite{mirror} suggests that the encoder outputs $\bm z=\mathcal{E}(\bm x)$ can be well approximated by a multivariate normal distribution. 
Moreover, we assume that the encoder $\mathcal{E}$ is Lipschitz continuous, \ie,
\begin{align}
||\hat{\bm{z}}-\bm{z}||=||\mathcal{E}(\hat{\bm{x}})-\mathcal{E}({\bm{x}})||\le L||\hat{\bm{x}}-\bm{x}||,
\end{align}
which is a standard assumption in the analysis of deep networks \cite{cisse2017parseval}. Under this condition, we can derive the following lower bound:
\begin{align}
&\hphantom{==}\text{IIE}(\mathcal{C}\circ\mathcal{E})\\
&=\mathop{\mathbb{E}}_{\bm{x}'\sim\mathcal{P}_{\bm{x}}}\left[\mathop{\min}_{\hat{\bm{x}}\in\mathbb{X}}\mathop{\mathbb{E}}_{\bm{x}\sim\mathcal{P}_{\bm{x}}}\left[\Vert\hat{\bm{x}}-\bm{x}\Vert^2_2\middle|\bm{W}\mathcal{E}(\bm{x})=\bm{W}\mathcal{E}(\bm{x}')\right]\right],\\
&\ge\mathop{\mathbb{E}}_{\bm{x}'\sim\mathcal{P}_{\bm{x}}}\left[\mathop{\min}_{\hat{\bm{x}}\in\mathbb{X}}\mathop{\mathbb{E}}_{\bm{x}\sim\mathcal{P}_{\bm{x}}}\left[\frac{\Vert\mathcal{E}(\hat{\bm{x}})-\mathcal{E}(\bm{x})\Vert^2_2}{L^2}\middle|\bm{W}\mathcal{E}(\bm{x})=\bm{W}\mathcal{E}(\bm{x}')\right]\right],\\
&=\mathop{\mathbb{E}}_{\bm{z}'\sim\mathcal{P}_{\mathcal{E}(\bm{x})}}\left[\mathop{\min}_{\hat{\bm{z}}\in\mathcal{E}(\mathbb{X})}\mathop{\mathbb{E}}_{\bm{z}\sim\mathcal{P}_{\mathcal{E}(\bm{x})}}\left[\frac{\Vert\hat{\bm{z}}-\bm{z}\Vert^2_2}{L^2}\middle|\bm{W}\bm{z}=\bm{W}\bm{z}'\right]\right],\\
&=\frac{1}{L^2}\mathop{\mathbb{E}}_{\bm{z}'\sim\mathcal{P}_{\mathcal{E}(\bm{x})}}\left[\mathop{\min}_{\hat{\bm{z}}\in\mathcal{E}(\mathbb{X})}\mathop{\mathbb{E}}_{\bm{z}\sim\mathcal{P}_{\mathcal{E}(\bm{x})}}\left[\Vert\hat{\bm{z}}-\bm{z}\Vert^2_2\middle|\bm{W}\bm{z}=\bm{W}\bm{z}'\right]\right],\\
&=\frac{1}{L^2}\text{IIE}(\mathcal{C}),\\
&\approx\frac{m-\text{rank}(\bm{W})}{L^2}.
\end{align}

This observation highlights that the ideal inversion error grows as the rank of the transformation matrix $\bm{W}$ decreases. In other words, lower-rank transformations inherently induce greater uncertainty in reconstruction, leading to a higher expected inversion error.
%
These findings emphasize the critical role of the transformation rank in trading off between model expressiveness and privacy. Specifically, lower-rank projections enhance privacy by increasing inversion uncertainty, although they may compromise utility due to loss of information. 

\begin{figure*}[t]
    \centering
    \includegraphics[width=.9\linewidth]{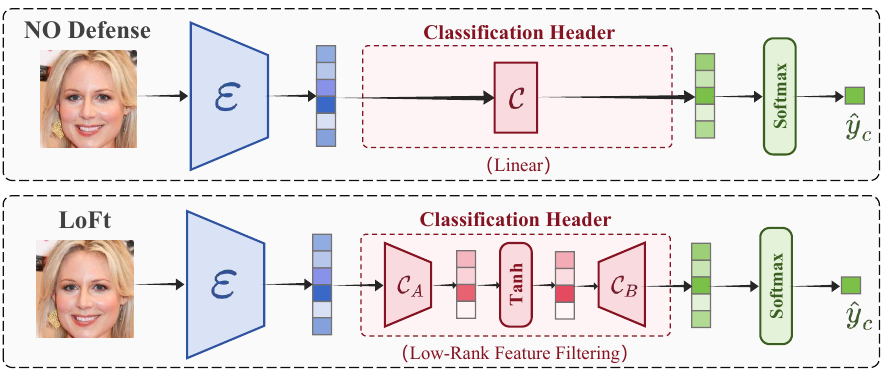}
    \caption{Overview of our LoFt defense strategy}
    \label{fig:method overview}
\end{figure*}

\section{Defense with Low-Rank Feature Filtering}
\label{sec:lowrank}


In the previous analysis, we find that a low-rank weight matrix can effectively enhance the reconstruction error. Motivated by this finding, we propose a \textbf{lo}w-rank \textbf{f}ea\textbf{t}ure filtering strategy (LoFt) by reducing the rank of weight matrix in the classification head.  To reduce the rank of the classification head’s weight matrix \( \bm{W} \in \mathbb{R}^{n \times m} \), we follow the low-rank approximation steps to reduce the rank to $r$:

\begin{enumerate}
    \item Apply Singular Value Decomposition (SVD): Decompose the matrix as \( \bm{W} = \bm{U} \bm{S} \bm{V}^T \), where \( \bm{S} \) is a diagonal matrix containing the singular values \( \lambda_1 \ge \lambda_2 \ge \dots \ge \lambda_{R} > 0 \), and \( R = \text{rank}(\bm{W}) \).
\item Truncate small singular values: Set the smallest \( R - r \) singular values to zero, \ie, assign \( \lambda_{r+1}, \lambda_{r+2}, \dots, \lambda_R \) to zero, resulting in a modified diagonal matrix \( \bm{S}' \). The retention ratio is defined as $\frac{\Sigma_{i=1}^r \lambda_i}{\Sigma_{j=1}^R \lambda_j}$.
\item Reconstruct a low-rank approximation: Compute the rank-\( r \) approximation of \( \bm{W} \) as \( \bm{W}' = \bm{U} \bm{S}' \bm{V}^T \). It retains the dominant components of the original matrix for the main task.
\item Update the classifier: Replace the weight matrix of \( \mathcal{C} \) with \( \bm{W}' \) to obtain a low-rank version of the classification head.
\end{enumerate}


\begin{figure}
    \centering
        \includegraphics[width=\linewidth]{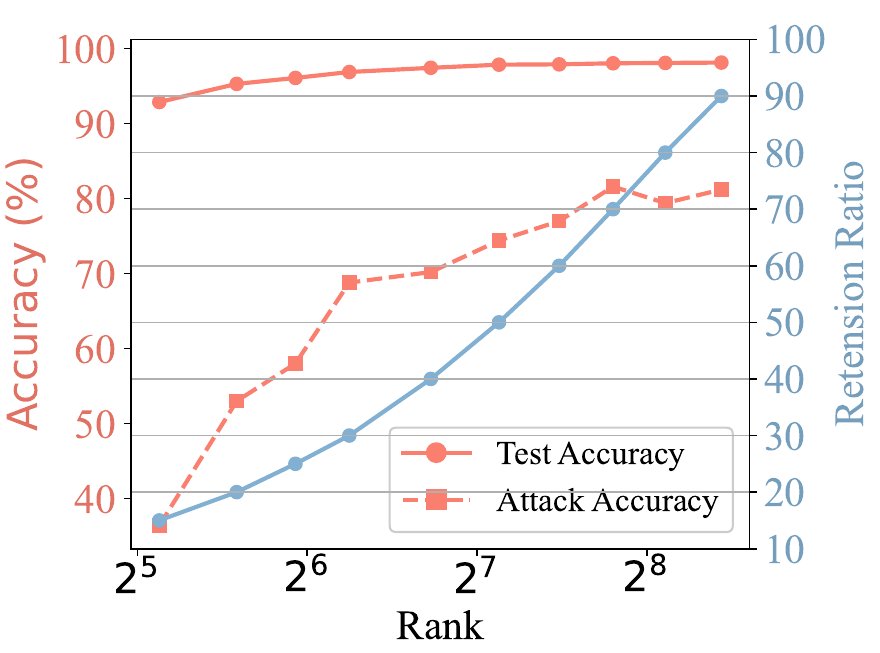}
        \caption{Experimental results with different rank $r$ via SVD. The red lines means test and IF attack accuracy with different compressed rank. The blue line indicates the retention ratio of the eigenvalue.}
        \label{fig: motivation conf acc}
\end{figure}

We apply the low-rank approximation to the classification head of an IR-$152$ model \cite{resnet} trained with the FaceScrub dataset \cite{facescrub} and evaluate their classification accuracy on the test dataset. Then we apply an advanced attack called IF \cite{ifgmi} to reconstruct the training data. The experimental details are provided in Sec. \ref{sec: experiment}. The results presented in Fig. \ref{fig: motivation conf acc} show that with the decrease of rank, the test accuracy decrease slowly, which means that the model only requires a few features to complete the main tasks. However, the attack accuracy drop sharply, which means that the low-rank approximation strategy removes a lot of redundant information that is not related to the main task but is essential to the attacks.

However, there are still some task-relevant information been removed, resulting that the test accuracy is less than $93\%$ when $r=35$. At the same time, some privacy-relevant information contains in the retained part, and the attack accuracy is not low enough. To address this issue, we continuously fine-tune the model with rank $r$ unchanged. However, directly optimize the weight matrix while keeping the rank unchanged is intractable. To address it,we further decompose $\mathcal{C}$ into two linear layers:  $\mathcal{C}= \mathcal{C}_B \circ \mathcal{C}_A$, with correspond matrix satisfies $\bm{W}^{n\times m}=\bm{W}^{n\times r}_B\bm{W}^{r\times m}_A$.

From another perspective, the first layer $\mathcal{C}_A$ compresses the $m$-dimensional feature into $r$ dimensions, where $r\ll \min\{m, n\}$. The second layer $\mathcal{C}_B$ then uses this compressed representation to make the classification predictions. In this case, reducing the rank of the weight matrix can be translated into limiting the dimensionality of the intermediate feature representation. The model is constrained to rely on limited features for downstream tasks, forcing it to filter out irrelevant information and focus only on what is essential for classification.


Moreover, \cite{relu} emphasize that certain non-linear activation functions, such as $\mathrm{Sigmoid}$ and $\mathrm{Tanh}$, are susceptible to the gradient vanishing problem. To further hinder the attacker's optimization process, we propose introducing a non-linear activation function after the compressed features, specifically positioned between the two linear layers of the classification head. The back-propagation characteristics of various activation functions suggest that the $\mathrm{Tanh}$ function is particularly prone to causing the gradient vanishing problem, which will be analyzed in detail in Appendix \ref{appx:backward function}.
As a result, we choose $\mathrm{Tanh}$ as the activation function in our approach.

\def\toycolorbarwidth{0.0365\textwidth}
\def\toywidth{0.2\textwidth}

\begin{figure}[!htbp]
    \centering
    \begin{subfigure}[b]{\toywidth}
        \centering
        \includegraphics[width=\textwidth]{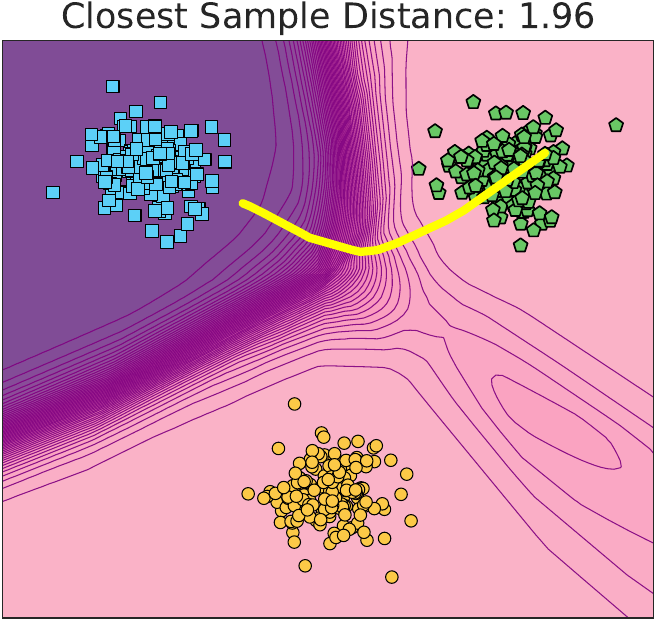}
        \caption{NO Defense}
        \label{fig:toyno}
    \end{subfigure}
    \begin{subfigure}[b]{\toywidth}
        \centering
        \includegraphics[width=\textwidth]{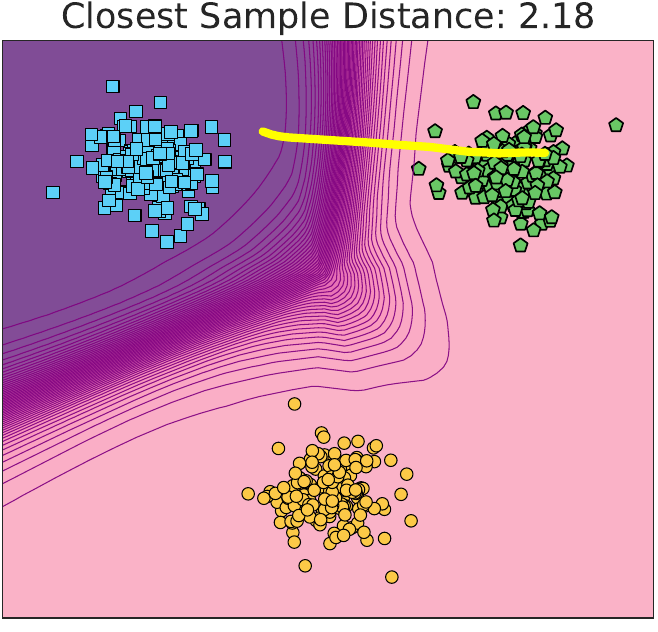}
        \caption{Tanh Activation}
        \label{fig:toytanh}
    \end{subfigure}
    \begin{subfigure}[b]{\toycolorbarwidth}
        \centering
        \includegraphics[width=\textwidth]{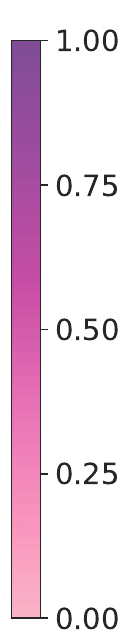}
        \caption*{}
    \end{subfigure}
    \begin{subfigure}[b]{\toywidth}
        \centering
        \includegraphics[width=\textwidth]{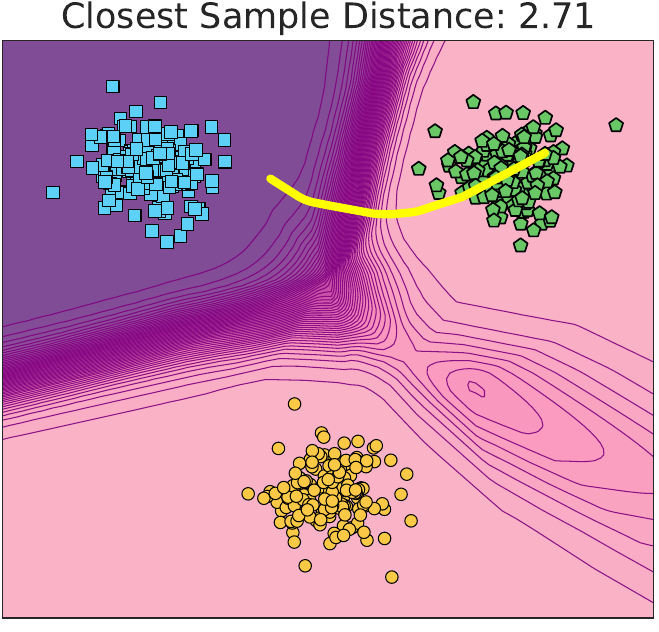}
        \caption{LoFt without Activation}
        \label{fig:toyneck}
    \end{subfigure}
    \begin{subfigure}[b]{\toywidth}
        \centering
        \includegraphics[width=\textwidth]{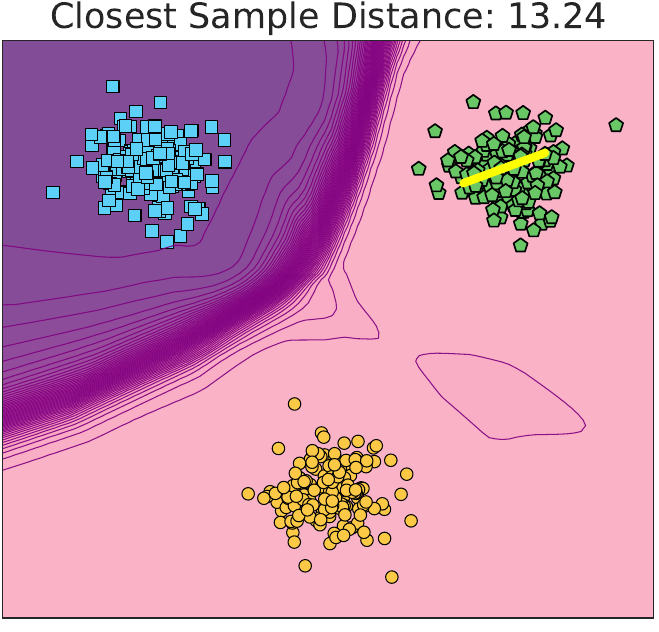}
        \caption{LoFt with Tanh}
        \label{fig:toynecktanh}
    \end{subfigure}
    \begin{subfigure}[b]{\toycolorbarwidth}
        \centering
        \includegraphics[width=\textwidth]{figure/toy3/colorbar.pdf}
        \caption*{}
    \end{subfigure}
    
    \caption{Simple MIA on a 2D toy dataset with three classes against different model architectures.
}
    \label{bigfig:toy fc}
\end{figure}

\begin{table*}[!ht]
    \setlength{\tabcolsep}{5pt}
    \normalsize
    \centering
    \caption{IF and PPDG attack results against IR-$152$ models trained on FaceScrub dataset in the low-resolution scenario.}
    \label{tab:low ffhq facescrub if plg}
    \begin{threeparttable} 
    \resizebox{\linewidth}{!}{
    \begin{tabular}{cccccccccc}
        \toprule
 \multirow{2}{*}{\textbf{Method}} 
 &  \multirow{2}{*}{\textbf{Test Acc}} 
 & \multicolumn{4}{c}{IF} &\multicolumn{4}{c}{PPDG} \\
 \cmidrule(lr){3-6}\cmidrule(lr){7-10} & & 
 $\downarrow$ \textbf{Acc@1} & $\downarrow$ \textbf{Acc@5} & $\uparrow$ \textbf{$\delta_{eval}$} & $\uparrow$ \textbf{$\delta_{face}$} & $\downarrow$ \textbf{Acc@1} & $\downarrow$ \textbf{Acc@5} & $\uparrow$ \textbf{$\delta_{eval}$} & $\uparrow$ \textbf{$\delta_{face}$}  \\ \midrule


\textbf{NO Defense} & $98.2$ & $82.2_{\pm 3.9}$ & $88.4_{\pm 2.4}$ & $1377_{\pm 414}$ & $0.70_{\pm 0.22}$ & $93.6_{\pm 14.9}$ & $97.8_{\pm 10.5}$ & $325_{\pm 55}$ & $0.60_{\pm 0.12}$ \\
\textbf{MID} & $97.0$ & $78.8_{\pm 4.9}$ & $87.0_{\pm 2.6}$ & $1479_{\pm 349}$ & $0.75_{\pm 0.16}$ & $95.6_{\pm 11.9}$ & $99.0_{\pm 5.2}$ & $306_{\pm 51}$ & $0.58_{\pm 0.12}$ \\
\textbf{BiDO} & $95.2$ & $51.0_{\pm 3.2}$ & $71.0_{\pm 3.5}$ & $1748_{\pm 374}$ & $0.81_{\pm 0.16}$ & $83.2_{\pm 23.1}$ & $95.4_{\pm 11.3}$ & $373_{\pm 55}$ & $0.71_{\pm 0.12}$ \\
\textbf{LS} & $97.3$ & $81.8_{\pm 1.9}$ & $87.6_{\pm 1.6}$ & $1392_{\pm 433}$ & $0.74_{\pm 0.27}$ & $94.6_{\pm 14.7}$ & $99.0_{\pm 6.6}$ & $326_{\pm 47}$ & $0.61_{\pm 0.10}$ \\
\textbf{TL} & $95.4$ & $42.0_{\pm 3.7}$ & $55.6_{\pm 1.4}$ & $1903_{\pm 467}$ & $0.94_{\pm 0.25}$ & $98.4_{\pm 5.4}$ & $99.8_{\pm 2.0}$ & $283_{\pm 36}$ & $0.51_{\pm 0.08}$ \\
\textbf{RoLSS} & $97.4$ & $56.6_{\pm 4.6}$ & $72.8_{\pm 4.0}$ & $1692_{\pm 404}$ & $0.84_{\pm 0.19}$ & $80.0_{\pm 23.0}$ & $92.6_{\pm 14.6}$ & $380_{\pm 63}$ & $0.73_{\pm 0.15}$ \\
\textbf{Trap} & $96.3$ & $26.8_{\pm 0.7}$ & $40.2_{\pm 2.8}$ & $2101_{\pm 462}$ & $1.11_{\pm 0.29}$ & $79.8_{\pm 23.9}$ & $90.0_{\pm 17.5}$ & $373_{\pm 78}$ & $0.70_{\pm 0.17}$ \\
\textbf{LoFt (ours)} & $97.1$ & $\mathbf{13.0}_{\pm 3.4}$ & $\mathbf{27.0}_{\pm 2.4}$ & $\mathbf{2299}_{\pm 403}$ & $\mathbf{1.29}_{\pm 0.29}$ & $\mathbf{38.4}_{\pm 30.0}$ & $\mathbf{61.2}_{\pm 29.8}$ & $\mathbf{500}_{\pm 87}$ & $\mathbf{1.01}_{\pm 0.22}$ \\

\bottomrule
    \end{tabular}
    }
    \end{threeparttable}
\end{table*}

In total, the complete schematic diagram of our method is shown in Fig. \ref{fig:method overview}. Following \cite{ls}, we present a simple toy example to show how our proposed low-rank feature filtering (LoFt) affect model privacy. The example uses a two-dimensional dataset with three classes: blue squares, green pentagons, and orange circles. The background color indicates the
models’ prediction confidence, and the yellow lines show the intermediate optimization steps of the
attack. The optimization starts from a random position, here a sample from the green pentagon class, and tries to reconstruct a sample from the blue square class. This optimization runs for $2500$ steps, and its goal is to reveal the features of the target class, which are just the coordinates of the training samples in this case.

We train a three-layer neural network with a hidden dimension of 20 and ReLU activation. This is our baseline model without any defense. 
Then, we create another model by replacing the last ReLU activation with a Tanh activation. We also create LoFt models with a dimension of $2$, both with and without the Tanh activation, following the method in Sec. \ref{sec:lowrank}. 
 Fig. \ref{bigfig:toy fc} shows that the models with LoFt (Figs. \ref{fig:toyneck} and \ref{fig:toynecktanh}) tend to give either very low or very high confidence predictions.
Without LoFt, the models, with tanh (Fig. \ref{fig:toytanh}) or without it (Fig. \ref{fig:toyno}), produce similar reconstructed $l_2$ distances: $1.96$ and $2.18$, respectively. 
But with LoFt, the attacker’s optimization results are farther from the target sample, as seen in Fig. \ref{fig:toyneck}, where the distance increases to $2.71$.
Additionally, using Tanh activation in the feature model cause gradient vanishing, as shown in Fig. \ref{fig:toynecktanh}. This make it much harder for the attacker to optimize, resulting in a final distance of $13.24$ away from the target sample. This indicates a significant reduction in the effectiveness of the attack.

\section{Experiment}
\label{sec: experiment}

\subsection{Experimental Protocol}


\paragraph{Datasets.} Following previous works \cite{gmi,ppa}, we focus on the facial recognition task on private datasets $\mathcal{D}_{\texttt{pri}}$, including FaceScrub \cite{facescrub} and CelebA \cite{celeba}. 
The FaceScrub dataset comprises $530$ unique identities, with $265$ actors and $265$ actresses. The CelebA dataset includes a much larger set of $10,177$ distinct identities, and only top-$1000$ identities with the most samples will be used \cite{gmi}. 
The images are cropped and resized to the resolution of $64\times 64$ and $224\times 224$ in low- and high-resolution scenarios, respectively. 

\paragraph{Models.} We trained classifiers for multiple architectures, including convolution-based IR-$152$ \cite{resnet}
, ResNet-$152$ \cite{resnet} and FaceNet-$112$ \cite{facenet112}, as well as transformer-based ViT-B/16 \cite{vit}, Swin-v2 \cite{swinv2} and MaxViT \cite{maxvit}. 
Among them, FaceNet-$112$  and MaxViT are used as evaluation models for the low- and high-resolution cases, respectively. 
%
For each remaining model architecture, we train with different defense algorithms as target models. 
Target models are trained on the private dataset like FaceScrub or CelebA.
More details are provided in Appendix \ref{appx:classifier training}.

\paragraph{Attacks.} We apply various attack methods in our experiments. Recent MIAs are primarily divided into two main directions of development. The first direction explores the prior knowledge of well-structured pre-trained GANs, such as GMI \cite{gmi},  Mirror \cite{mirror}, PPA \cite{ppa}, and IF \cite{ifgmi}. The second focuses on fully leveraging the target model to train target-specific GANs or surrogate classifiers, including KED \cite{ked}, LOMMA \cite{lomma}, PLG \cite{plg}, and PPDG \cite{ppdg}.  In the main paper, we experiment with the most advanced algorithms from each of these two directions, \textit{i.e.}, IF and PPDG. The results of other attacks are available in Appendix \ref{appx:additional result}. More details of the attacks are shown in Appendix \ref{appx:attacks}.
%
%


\paragraph{Metrics.}
Following prior work \cite{ls,ifgmi}, we adopt several quantitative metrics to evaluate the effectiveness of different defense mechanisms, primarily including attack accuracy and feature distance. For all reported results, we provide both the mean and standard deviation to capture attack performance consistency across multiple runs.
We also analyze some other aspects of model robustness in Appendix \ref{appx:other robustness}, including the knowledge extraction scores \cite{ls}, adversarial robustness \cite{fgsm,pgd,bim,onepixel} and backdoor robustness \cite{badnets,blended}. 

\begin{itemize}
\item  \textit{Attack Accuracy.}
We use the evaluation model to predict the labels on reconstructed samples and compute the top-$1$ and top-$5$ accuracy for target classes, denoted as $\textbf{Acc@1}$ and $\textbf{Acc@5}$, respectively. Higher attack accuracy indicates a greater leakage of private information \cite{gmi}.
\item \textit{Feature Distance.} 
Features are defined as the output from the model's penultimate layer. For each reconstructed sample, we calculate the $l_2$ feature distance to the nearest private training sample. The final metric is obtained by averaging these distances across all reconstructed samples. 
The feature distances are evaluated using the evaluation model and a FaceNet \cite{schroff2015facenet} trained on a large VGGFace2 dataset \cite{vggface2}, denoted as $\delta_{eval}$ and $\delta_{face}$.
\end{itemize}

\paragraph{Defense baselines.} We compare our method with $6$ SOTA defense algorithms, including MID \cite{mid}, BiDO \cite{bido}, LS \cite{ls}, TL \cite{tl}, RoLSS \cite{rolss} and Trap \cite{trap}. To ensure a fair comparison among the different defense algorithms, we carefully adjust the hyperparameters governing defense strength, maintaining nearly identical classification accuracy on the test split of the private dataset for each model, denoted as \textbf{Test Acc}. The detailed implementation of the defense methods is provided in the Appendix \ref{appx:classifier training}.

\begin{table*}[!t]
    \setlength{\tabcolsep}{5pt}
    \normalsize
    \centering
    \caption{IF and PPDG attack results against ResNet-$152$ models trained on FaceScrub dataset in the high-resolution scenario.}
    \label{tab:high ffhq facescrub if plg}
    \begin{threeparttable} 
    \resizebox{\linewidth}{!}{
    \begin{tabular}{ccccccccccc}
        \toprule
 \multirow{2}{*}{\textbf{Pretrained}} 
 & \multirow{2}{*}{\textbf{Method}} 
 &  \multirow{2}{*}{\textbf{Test Acc}} 
 & \multicolumn{4}{c}{IF} &\multicolumn{4}{c}{PPDG} \\
 \cmidrule(lr){4-7}\cmidrule(lr){8-11} & & &
  $\downarrow$ \textbf{Acc@1} & $\downarrow$ \textbf{Acc@5} & $\uparrow$ \textbf{$\delta_{eval}$} & $\uparrow$ \textbf{$\delta_{face}$} & $\downarrow$ \textbf{Acc@1} & $\downarrow$ \textbf{Acc@5} & $\uparrow$ \textbf{$\delta_{eval}$} & $\uparrow$ \textbf{$\delta_{face}$} \\\midrule

\multirow{7}{*}{\ding{55}} & \textbf{NO Defense} & $92.2$ & $85.8_{\pm 2.1}$ & $96.8_{\pm 1.6}$ & $377_{\pm 59}$ & $0.71_{\pm 0.14}$ & $83.4_{\pm 25.5}$ & $93.0_{\pm 16.1}$ & $382_{\pm 67}$ & $0.71_{\pm 0.14}$ \\
 & \textbf{MID} & $88.2$ & $81.2_{\pm 4.3}$ & $90.0_{\pm 2.8}$ & $392_{\pm 65}$ & $0.77_{\pm 0.15}$ & $78.6_{\pm 25.8}$ & $92.6_{\pm 14.0}$ & $395_{\pm 66}$ & $0.76_{\pm 0.15}$ \\
 & \textbf{BiDO} & $88.6$ & $79.2_{\pm 3.3}$ & $92.4_{\pm 1.2}$ & $392_{\pm 65}$ & $0.73_{\pm 0.15}$ & $71.2_{\pm 30.8}$ & $92.6_{\pm 16.4}$ & $396_{\pm 63}$ & $0.73_{\pm 0.14}$ \\
 & \textbf{LS} & $88.8$ & $29.6_{\pm 5.0}$ & $57.4_{\pm 3.0}$ & $508_{\pm 67}$ & $1.01_{\pm 0.17}$ & $27.4_{\pm 27.3}$ & $55.8_{\pm 35.2}$ & $521_{\pm 73}$ & $1.06_{\pm 0.18}$ \\
 & \textbf{TL} & $88.4$ & $64.6_{\pm 1.9}$ & $85.0_{\pm 3.0}$ & $420_{\pm 59}$ & $0.81_{\pm 0.14}$ & $47.0_{\pm 33.7}$ & $72.2_{\pm 30.7}$ & $459_{\pm 75}$ & $0.91_{\pm 0.17}$ \\
 & \textbf{RoLSS} & $88.7$ & $52.6_{\pm 3.9}$ & $79.4_{\pm 2.1}$ & $444_{\pm 60}$ & $0.88_{\pm 0.14}$ & $54.4_{\pm 34.9}$ & $82.2_{\pm 24.3}$ & $436_{\pm 71}$ & $0.86_{\pm 0.16}$ \\
 & \textbf{Trap} & $85.7$ & $72.0_{\pm 1.9}$ & $91.4_{\pm 1.2}$ & $276_{\pm 61}$ & $0.73_{\pm 0.14}$ & $77.2_{\pm 29.8}$ & $93.6_{\pm 16.0}$ & $384_{\pm 69}$ & $0.74_{\pm 0.15}$ \\
 & \textbf{LoFt (ours)} & $89.4$ & $\mathbf{28.6}_{\pm 4.2}$ & $\mathbf{52.8}_{\pm 1.7}$ & $\mathbf{519}_{\pm 86}$ & $\mathbf{1.06}_{\pm 0.21}$ & $\mathbf{19.2}_{\pm 23.0}$ & $\mathbf{40.8}_{\pm 30.6}$ & $\mathbf{560}_{\pm 91}$ & $\mathbf{1.16}_{\pm 0.22}$ \\
\midrule
\multirow{7}{*}{$\checkmark$} & \textbf{NO Defense} & $98.5$ & $99.2_{\pm 0.7}$ & $99.6_{\pm 0.5}$ & $286_{\pm 47}$ & $0.52_{\pm 0.12}$ & $98.6_{\pm 7.1}$ & $99.8_{\pm 2.0}$ & $293_{\pm 53}$ & $0.52_{\pm 0.12}$ \\
 & \textbf{MID} & $96.8$ & $99.8_{\pm 0.4}$ & $100.0_{\pm 0.0}$ & $251_{\pm 37}$ & $0.46_{\pm 0.09}$ & $99.4_{\pm 3.4}$ & $100.0_{\pm 0.0}$ & $258_{\pm 41}$ & $0.47_{\pm 0.09}$ \\
 & \textbf{BiDO} & $96.3$ & $98.4_{\pm 0.8}$ & $99.8_{\pm 0.4}$ & $302_{\pm 45}$ & $0.55_{\pm 0.10}$ & $92.4_{\pm 17.4}$ & $98.4_{\pm 6.7}$ & $331_{\pm 61}$ & $0.60_{\pm 0.13}$ \\
 & \textbf{LS} & $96.4$ & $99.0_{\pm 0.6}$ & $100.0_{\pm 0.0}$ & $295_{\pm 39}$ & $0.55_{\pm 0.10}$ & $98.2_{\pm 9.8}$ & $100.0_{\pm 0.0}$ & $295_{\pm 46}$ & $0.53_{\pm 0.10}$ \\
 & \textbf{TL} & $96.6$ & $100.0_{\pm 0.0}$ & $100.0_{\pm 0.0}$ & $235_{\pm 37}$ & $0.42_{\pm 0.08}$ & $99.8_{\pm 2.0}$ & $100.0_{\pm 0.0}$ & $249_{\pm 41}$ & $0.43_{\pm 0.09}$ \\
 & \textbf{RoLSS} & $96.0$ & $93.8_{\pm 2.6}$ & $99.2_{\pm 0.4}$ & $352_{\pm 60}$ & $0.65_{\pm 0.12}$ & $92.8_{\pm 15.6}$ & $98.2_{\pm 9.4}$ & $354_{\pm 62}$ & $0.67_{\pm 0.15}$ \\
 & \textbf{Trap} & $96.7$ & $96.4_{\pm 1.2}$ & $97.8_{\pm 1.2}$ & $303_{\pm 68}$ & $0.54_{\pm 0.14}$ & $96.8_{\pm 10.5}$ & $98.6_{\pm 8.1}$ & $310_{\pm 63}$ & $0.55_{\pm 0.14}$ \\
 & \textbf{LoFt (ours)} & $96.7$ & $\mathbf{64.8}_{\pm 4.5}$ & $\mathbf{83.2}_{\pm 2.3}$ & $\mathbf{431}_{\pm 74}$ & $\mathbf{0.86}_{\pm 0.19}$ & $\mathbf{55.8}_{\pm 35.0}$ & $\mathbf{79.0}_{\pm 28.9}$ & $\mathbf{444}_{\pm 90}$ & $\mathbf{0.88}_{\pm 0.22}$ \\

\bottomrule
    \end{tabular}
    }
    \end{threeparttable}
\end{table*}

\subsection{Comparison with Previous Defense Methods}

%

\textbf{Low-resolution scenarios.} We provide quantitative attack results against IR-$152$ trained on the FaceScrub dataset in Table \ref{tab:low ffhq facescrub if plg}. We can observe that our method achieves significant improvements over previous defense methods.
Especially in terms of attack accuracy, compared to the most advanced baseline in this scenario, \ie, Trap, the attack accuracy is reduced by more than half, while maintaining higher test accuracy. We also conduct additional experiments with other attack methods. The results presented in Appendix \ref{appx:additional result} show that our approach demonstrates effective robustness across a wide range of model inversion attacks.

%
\textbf{High-resolution scenarios.} In these scenarios, previous studies \cite{ppa,ls,ifgmi} have utilized target models that are directly trained on the private dataset.
However, this approach results in models with insufficient accuracy, failing to accurately represent real-world conditions \cite{sevastopolskiy2023boost}.
To bridge this gap, we pre-train the models on large-scale public face datasets. This pre-training process allows the models to achieve a test accuracy of over $96\%$ on the FaceScrub dataset, showing strong performance on their main tasks.
We conduct experiments under two conditions, with and without pre-training. The attack results are presented in Table \ref{tab:high ffhq facescrub if plg}. Our method demonstrates superior defense performance.
When models show low accuracy with direct training, consistent with previous studies, recent defense methods such as LS and TL demonstrate improved defense capabilities. However, our approach still achieves the best defensive performance.
Notably, in scenarios with high classification accuracy, previous methods show significant difficulty in defending against the most advanced attacks.
In contrast, our method reduces the attack accuracy of IF and PPDG by $34.4\%$ and $42.8\%$, respectively, and increases the feature distance between the reconstructed and private images to $1.5$ to $2$ times compared to the case without defense.

\textbf{Visualization results.} The visualization results of the reconstructed images from IF attacks under various defense methods are shown in Fig. \ref{fig:high ffhq facescrub msceleb}. Compared to previous approaches, our defense strategy significantly increases the disparity between the reconstructed images and the original private images.

{

\def\cvresultwidth{0.11\linewidth}
\def\cvresultinnerwidth{0.98\linewidth}

\centering
\begin{figure}[!htbp]
\centering
\setlength{\tabcolsep}{1pt}
  \normalsize
\centering
\begin{subfigure}{0.99\linewidth}
    \begin{minipage}[t]{\cvresultwidth}
    \centering
    \includegraphics[width=\cvresultinnerwidth]{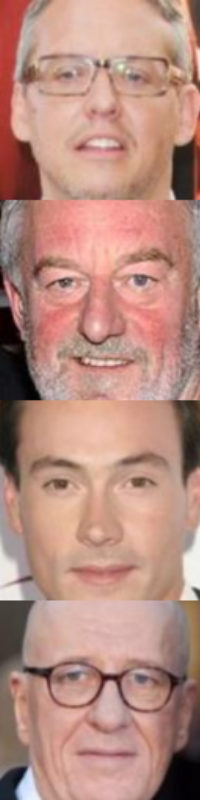}
    \centering
    \caption*{\centering\textbf{\footnotesize{Private Image}}}
    \end{minipage}%
    \begin{minipage}[t]{\cvresultwidth}
    \centering
    \includegraphics[width=\cvresultinnerwidth]{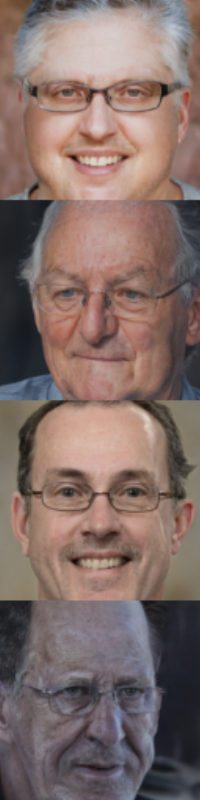}
    \centering
    \caption*{\centering\textbf{\footnotesize{NO\\Defense}}}
    \end{minipage}%
    \begin{minipage}[t]{\cvresultwidth}
    \centering
    \includegraphics[width=\cvresultinnerwidth]{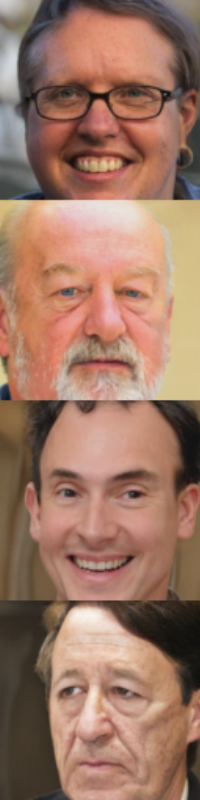}
    \centering
    \caption*{\centering\textbf{\footnotesize{MID}}}
    \end{minipage}%
    \begin{minipage}[t]{\cvresultwidth}
    \centering
    \includegraphics[width=\cvresultinnerwidth]{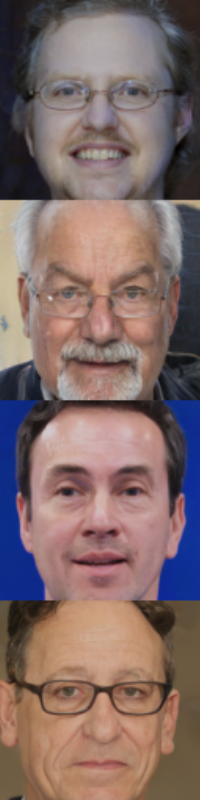}
    \centering
    \caption*{\centering\textbf{\footnotesize{BiDO}}}
    \end{minipage}%
    \begin{minipage}[t]{\cvresultwidth}
    \centering
    \includegraphics[width=\cvresultinnerwidth]{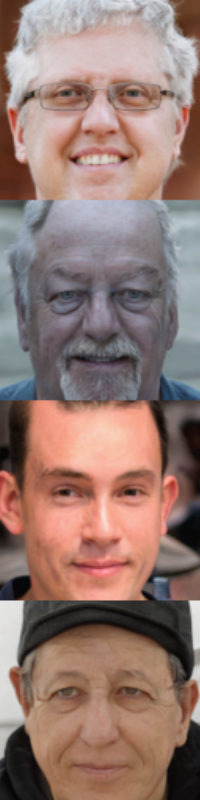}
    \centering
    \caption*{\centering\textbf{\footnotesize{LS}}}
    \end{minipage}%
    \begin{minipage}[t]{\cvresultwidth}
    \centering
    \includegraphics[width=\cvresultinnerwidth]{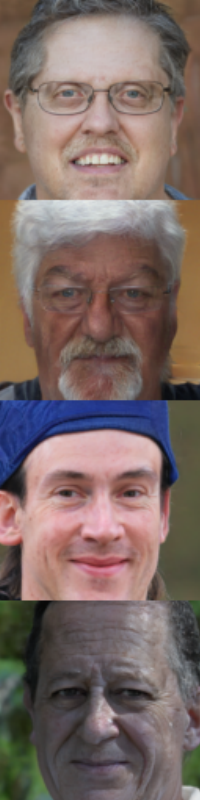}
    \centering
    \caption*{\centering\textbf{\footnotesize{TL}}}
    \end{minipage}%
    \begin{minipage}[t]{\cvresultwidth}
    \centering
    \includegraphics[width=\cvresultinnerwidth]{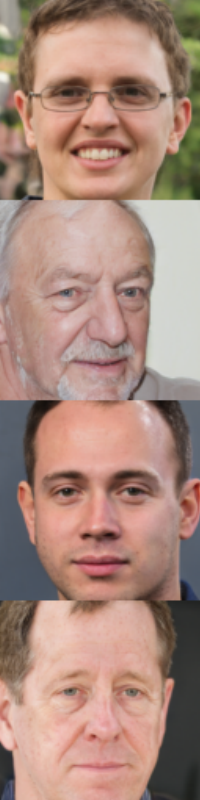}
    \centering
    \caption*{\centering\textbf{\footnotesize{RoLSS}}}
    \end{minipage}%
    \begin{minipage}[t]{\cvresultwidth}
    \centering
    \includegraphics[width=\cvresultinnerwidth]{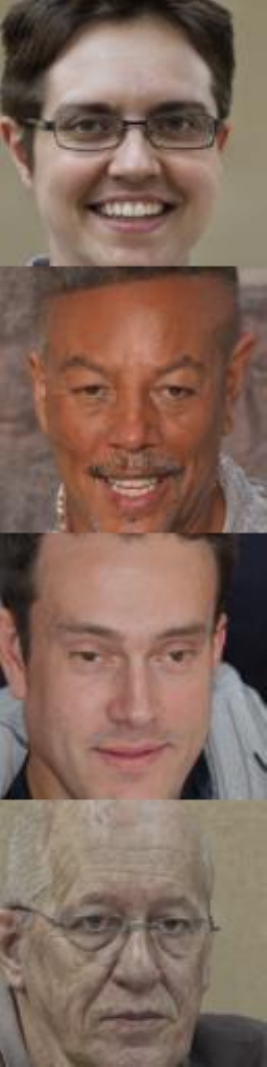}
    \centering
    \caption*{\centering\textbf{\footnotesize{Trap}}}
    \end{minipage}%
    \begin{minipage}[t]{\cvresultwidth}
    \centering
    \includegraphics[width=\cvresultinnerwidth]{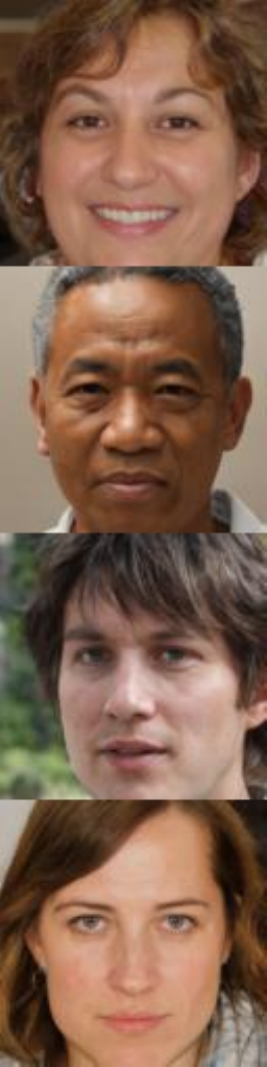}
    \centering
    \caption*{\centering\textbf{\footnotesize{LoFt\\(ours)}}}
    \end{minipage}%
\end{subfigure}
\caption{Visual comparison of IF attacks against ResNet-$152$ under different defense strategies.}
\label{fig:high ffhq facescrub msceleb}
\end{figure}

}



\begin{table*}[!htbp]
    \setlength{\tabcolsep}{5pt}
    \normalsize
    \centering
    \caption{The influence of different non-linear activation functions on IF attacks. }
    \label{tab:grad result}
    \begin{threeparttable} 
    \resizebox{\linewidth}{!}{
    \begin{tabular}{ccccccccccc}
        \toprule
 \multirow{2}{*}{\textbf{Activation}} 
 &   \multicolumn{5}{c}{Low Resolution} &\multicolumn{5}{c}{High Resolution} \\
 \cmidrule(lr){2-6}\cmidrule(lr){7-11} & \textbf{Test Acc}& 
 $\downarrow$ \textbf{Acc@1} & $\downarrow$ \textbf{Acc@5} & $\uparrow$ \textbf{$\delta_{eval}$} & $\uparrow$ \textbf{$\delta_{face}$} & \textbf{Test Acc} & $\downarrow$ \textbf{Acc@1} & $\downarrow$ \textbf{Acc@5} & $\uparrow$ \textbf{$\delta_{eval}$} & $\uparrow$ \textbf{$\delta_{face}$}  \\ \midrule

\textbf{Identity} & $94.1$ & $12.6_{\pm 4.3}$ & $24.2_{\pm 4.5}$ & $2304_{\pm 405}$ & $1.30_{\pm 0.27}$ & $96.7$ & $82.0_{\pm 1.7}$ & $91.6_{\pm 2.2}$ & $412_{\pm 64}$ & $0.78_{\pm 0.15}$ \\
\textbf{ReLU} & $92.7$ & $14.2_{\pm 2.3}$ & $21.8_{\pm 2.5}$ & $2389_{\pm 402}$ & $1.29_{\pm 0.27}$ & $96.1$ & $81.6_{\pm 2.6}$ & $93.2_{\pm 0.7}$ & $402_{\pm 60}$ & $0.79_{\pm 0.14}$ \\
\textbf{Sigmoid} & $93.2$ & $7.4_{\pm 1.6}$ & $16.2_{\pm 2.5}$ & $2405_{\pm 363}$ & $1.39_{\pm 0.28}$ & $\mathbf{96.8}$ & $67.6_{\pm 2.9}$ & $84.8_{\pm 3.2}$ & $417_{\pm 62}$ & $0.83_{\pm 0.16}$ \\
\textbf{Tanh} & $\mathbf{96.1}$ & $\mathbf{4.4}_{\pm 1.5}$ & $\mathbf{11.8}_{\pm 4.1}$ & $\mathbf{2475}_{\pm 330}$ & $\mathbf{1.40}_{\pm 0.26}$ & $96.6$ & $\mathbf{64.6}_{\pm 7.3}$ & $\mathbf{77.4}_{\pm 4.2}$ & $\mathbf{441}_{\pm 80}$ & $\mathbf{0.88}_{\pm 0.20}$ \\

\bottomrule
    \end{tabular}
    }
    \end{threeparttable}
\end{table*}

\begin{figure}
    \centering
        \includegraphics[width=.9\linewidth]{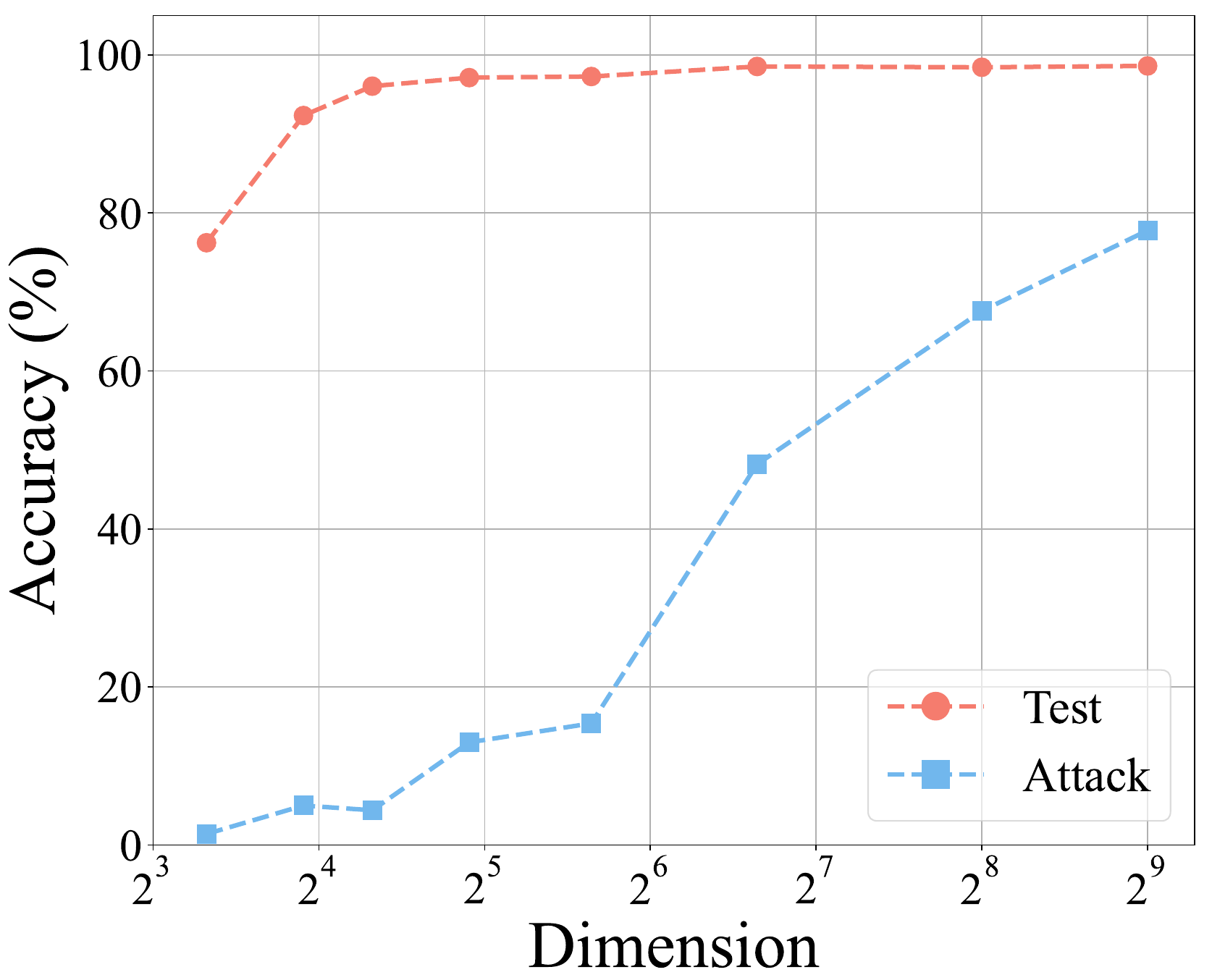}
        \caption{Test accuracy and IF attack accuracy with different compressed dimension.}
        \label{fig:rankacc}
\end{figure}


\begin{table}[!t]
    \setlength{\tabcolsep}{5pt}
    \normalsize
    \centering
    \caption{IF attack results against IR-$152$ models with different dimensions in low-resolution scenarios.}
    \label{tab:rank result}
    \vspace{5pt}
    \begin{threeparttable} 
    \resizebox{\linewidth}{!}{
    \begin{tabular}{cccccccccc}
        \toprule
{\textbf{Dimension}} 
 &  {\textbf{Test Acc}} &
 $\downarrow$ \textbf{Acc@1} & $\downarrow$ \textbf{Acc@5} & $\uparrow$ \textbf{$\delta_{eval}$} & $\uparrow$ \textbf{$\delta_{face}$}  \\ \midrule
\textbf{10} & $76.2$ & $1.4_{\pm 1.0}$ & $5.4_{\pm 2.7}$ & $2653_{\pm 298}$ & $1.49_{\pm 0.25}$ \\
\textbf{15} & $92.3$ & $5.0_{\pm 1.4}$ & $9.6_{\pm 1.4}$ & $2528_{\pm 313}$ & $1.41_{\pm 0.24}$ \\
\textbf{20} & $96.1$ & $4.4_{\pm 1.5}$ & $11.8_{\pm 4.1}$ & $2475_{\pm 330}$ & $1.40_{\pm 0.26}$ \\
\textbf{30} & $97.1$ & $13.0_{\pm 3.4}$ & $27.0_{\pm 2.4}$ & $2299_{\pm 403}$ & $1.29_{\pm 0.29}$ \\
\textbf{50} & $97.3$ & $15.4_{\pm 0.8}$ & $30.8_{\pm 2.8}$ & $2219_{\pm 359}$ & $1.18_{\pm 0.23}$ \\
\textbf{100} & $98.6$ & $48.2_{\pm 2.9}$ & $66.4_{\pm 3.3}$ & $1806_{\pm 379}$ & $0.97_{\pm 0.25}$ \\
\textbf{256} & $98.4$ & $67.6_{\pm 2.6}$ & $79.8_{\pm 3.9}$ & $1669_{\pm 354}$ & $0.84_{\pm 0.20}$ \\
\textbf{512} & $98.6$ & $77.8_{\pm 3.5}$ & $86.8_{\pm 3.1}$ & $1519_{\pm 331}$ & $0.79_{\pm 0.20}$ \\
\bottomrule
    \end{tabular}
}
    \end{threeparttable}
\end{table}

\textbf{Comparison in more experimental settings.} In addition to the previously discussed experiments, we conduct further comparative studies under a wider range of experimental settings.
These additional settings include more attack methods, more private datasets such as CelebA, and more target classifier architectures, including transformer-based models like Swin-v2 and ViT-B/16.
The experimental results shown in Appendix \ref{appx:additional result} demonstrate that our defense method consistently achieves strong performance across a variety of settings.
%
%

\subsection{Ablation Study}




\textbf{The compressed dimension of the classification head.} To investigate the effect of the classification header's compressed dimension on MIAs, we train a series of IR-$152$ and ResNet-$152$ models with varying dimensions and evaluate their test accuracy.
Subsequently, we apply the IF attacks on these models and record the attack accuracy and distance metric results.
The results in low-resolution scenarios are provided in Table  \ref{tab:rank result} and Fig. \ref{fig:rankacc}. For high-resolution scenarios, the results for ResNet-$152$ models are available in Appendix \ref{appx:rank result high}. The findings reveal that a low dimension, such as $30$ in low-resolution scenarios, is sufficient for the model to effectively capture essential features necessary for the classification task, leading to strong overall performance. In this context, the model exhibits notable robustness against attacks, achieving a low attack accuracy of only  $13.0\%$. 
Moreover, with the increase in the dimension, such as from 100 to 512, there are only marginal gains in model performance; however, it significantly boosts the attacker’s effectiveness. Specifically, the attack accuracy increases from $48.2\%$ to $77.8\%$, raising serious concerns about privacy leakage. These observations indicate that adopting The LoFt strategy not only maintains high model performance, but also serves as an effective defense against MIAs. This dual benefit highlights the potential of LoFt approaches in ensuring both efficacy and security in model deployment.

%


\begin{figure}[!h]
        \centering
        \includegraphics[width=.7\linewidth]{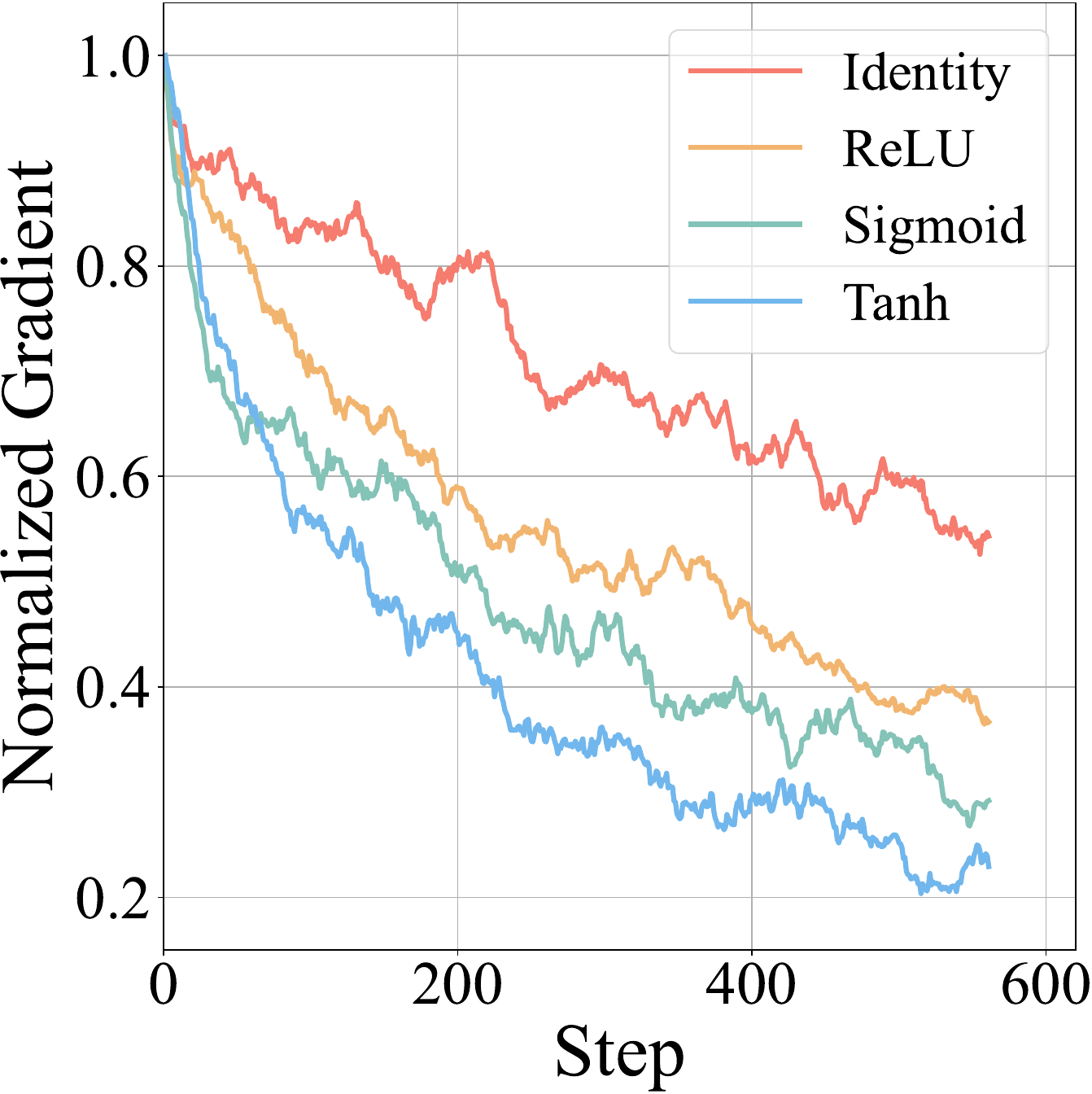}
        \caption{The trend of gradient magnitude in IF attacks.}
        \label{fig:grad}
\end{figure}

\textbf{Gradient vanishing of the non-linear activation.}
\label{para:nonlinear activation}
We begin by analyzing the gradient vanishing effect of different non-linear activation functions.
We apply IF attacks on ResNet-$152$ models in the high-resolution scenario, using different non-linear activation functions. During this process, we record the gradient magnitudes at each optimization step, which is displayed in Fig. \ref{fig:grad}. 
The results indicate a rapid decline in gradient magnitude when non-linear activation functions are applied to the model, with the $\mathrm{Tanh}$ activation function exhibiting the most significant decrease.
Table \ref{tab:grad result} presents the IF attack outcomes for each activation function in both low- and high-resolution scenarios. The results also reveal that the $\mathrm{Tanh}$ activation achieves an optimal balance between model performance and model inversion defense effectiveness.


\section{Conclusion}

Deep Neural Networks (DNNs) have tremendous potential across various domains. However, its applications must ensure robust privacy protections to prevent private data from leakage, which can be compromised by model inversion attacks (MIAs). 
Our theoretical and empirical investigations reveal that the higher rank of intermediate representations results in a higher risk of privacy leakage. Building on this insight, we propose a low-rank feature suppression strategy by constraining the dimension of the intermediate representations. 
Through extensive experimental validation, our defense method demonstrates state-of-the-art (SOTA) effectiveness across diverse scenarios. Notably, it excels at protecting models while maintaining high performance, surpassing prior defenses. We hope this work provides valuable insights for mitigating privacy leakage risks in deployed models and fosters a deeper understanding of defense mechanisms against MIAs.

\begin{acks}
This work is supported in part by the National Science Foundation for Distinguished Young Scholars of China under No. 62425201, and the National Natural Science Foundation of China under grant 62571298, 62576122, 62301189, and Shenzhen Science and Technology Program under Grant KJZD20240903103702004.
To Robert, for the bagels and explaining CMYK and color spaces.
\end{acks}



\appendix

\section{Summary of Previous Defense Methods}
\label{appx:defense summary}

In this section, we briefly summarize previous defense methods and introduce how they defend against MIAs.





\paragraph{Mutual Information Regularization based Defense (MID).} \cite{mid} enhance model robustness by reducing the mutual information between the model's input and output. The formulation of the loss is as follows:
\begin{equation}
    \mathcal{L}_{MID} = \mathcal{L}_{CE} +  \lambda I(X, \hat Y),
\end{equation}
where $\lambda$ is a hyperparameter and $I(X, \hat{Y})$ denotes the mutual information between inputs and outputs. However, this mutual information loss term cannot be calculated directly.
To overcome this, they apply a variational approach \cite{alemi2016deep} to approximate the mutual information loss term. 
In practice, this involves Gaussian sampling at the output of the penultimate layer.  The loss functions are as follows:
\begin{equation}
    \mathcal{L}_{MID}=\mathcal{L}_{CE} + \lambda (-\frac{1}{2}(1+\log \sigma^2 - \mu^2 - \sigma^2)),
\end{equation}
where $\lambda$ is a hyperparameter, $\mu$ and $\sigma$ are the output of the penultimate layer to execute the Gaussian sampling.

\paragraph{Bilateral Dependency Optimization (BiDO).} \cite{bido} propose to reduce the dependency $d(\cdot, \cdot)$ between the model inputs $X$ and the intermediate feature $Z$ and improve that between $Z$ and labels ${Y}$.  The loss function is:
\begin{equation}
    \mathcal{L}_{BiDO}=\mathcal{L}_{CE} + \lambda_{iz} \sum_{i=1}^M d(X, Z_i) - \lambda_{oz} \sum_{i=1}^M d(Z_i, Y),
\end{equation}
where $\lambda_{iz}$ and $\lambda_{oz}$ are hyperparameters. $Z_i, i\in[1,2,\dots M]$ represents different intermediate output by the model. In our experiment, we set $M=3$.
The dependency measure $d(\cdot, \cdot)$ can be represented by Constrained Covariance (COCO) \cite{coco} or the Hilbert-Schmidt Independence Criterion (HSIC) \cite{hsic}. According to the original paper, the HSIC criterion demonstrates superior performance in defense. Therefore, we utilize the HSIC criterion in our experiment.

\paragraph{Transfer Learning (TL).} \cite{tl} analyse the fisher information of each layers in classification and inversion tasks. Through the analysis of fisher information between the parameters and different tasks, they find that the previous layers in the model are more important for the model inversion task  and the last some layers makes more contributions to the classification task.
Therefore, to migrate the effect of MIAs, they pre-train the model on public datasets and freeze the parameters in previous layers when fine-tuning with private datasets.
%
The hyperparameter is the freezing ratio, denoted as $\lambda$.



\paragraph{Label Smoothing (LS).} 
\cite{ls} explore the effect of the label smoothing technique to MIAs. The loss function of LS is:
\begin{equation}
    \mathcal{L}_{LS} = 
    (1-\lambda)\mathcal{L}_{CE}(\mathbf{y}, \mathbf{\hat y}) 
    + \frac{\lambda}{C} \mathcal{L}_{CE}(\mathbf{1}, \mathbf{\hat y}),
\end{equation}
where $\lambda$ is the label smoothing factor, $\mathbf{y}$ is the label, $\mathbf{\hat y}$ is the model prediction and $\mathbf{1}$ denotes a vector of length $C$ with all entries set to 1. $C$ is the number of the classes. 
The factor $\lambda>0$ will facilitate the inversion attack, while  $\lambda<0$ will have a strong defensive effect.

\begin{figure}[!htbp]
    \centering
    \begin{subfigure}[b]{0.22\textwidth}
    \begin{minipage}[b]{\textwidth}
        \centering
        \includegraphics[width=\textwidth]{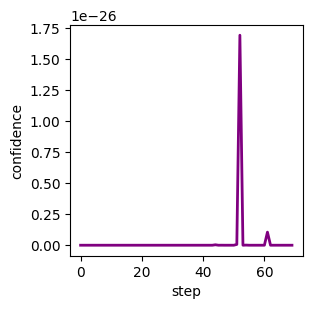}
        \caption{Not Pre-trained}
        \label{fig:lsconf nopre}
    \end{minipage}%
    \end{subfigure}
    \begin{subfigure}[b]{0.22\textwidth}
    \begin{minipage}[b]{\textwidth}
        \centering
        \includegraphics[width=\textwidth]{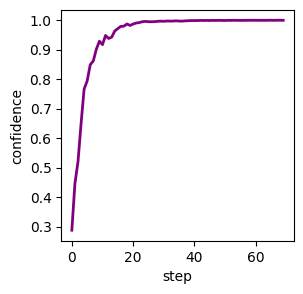}
    \end{minipage}%
        \caption{Pre-trained}
        \label{fig:lsconf pre}
    \end{subfigure}
    \caption{The trend of classifier confidence under IF attack across different pre-trained datasets $\mathcal{D}_\texttt{pre}$ with the LS defense method.}
    \label{fig:lsconf}
\end{figure}

LS with a negative label smoothing factor can effectively impede the optimization process of the attacks. To assess this, we select some challenging samples as target identities and apply the IF attacks on a ResNet-$152$ model pre-trained with various public datasets, recording the confidence levels predicted by the victim classifier at each optimization step.
The results are shown in Fig. \ref{fig:lsconf}. When the model is not pre-trained, we observe that the target confidence remains consistently low, indicating that the optimization process is significantly obstructed. 
However, when the model is pre-trained on a large public face dataset, the obstruction effect diminish, making it more challenging to defend against the attack in this scenario.
%
We hypothesize that directly training on the private dataset necessitates substantial adjustment of model parameters during training, allowing LS to serve as an effective defense. Conversely the adjustments after pre-training are minimal, which may reduce the effectiveness of LS in this context. 

\paragraph{Removal of Last Stage
Skip-Connection (RoLSS).} Skip connections were first proposed by \cite{resnet}, which can effectively solve the gradient vanishing problem on training deep neural network. However, \cite{rolss} found that this is also helpful in guiding the attacker's optimization process. Therefore, they decide to remove the skip connections of the last few layers of the model to hinder the attacker's optimization process. In our experiments, the hyperparameter $\lambda$ is the number of skip connections to be removed.

\paragraph{Trapdoor-based Defense (Trap).}
Trapdoor-based Defense \cite{trap} propose to inject triggers into the model, so that the MI attacks will follow the "shortcut" to reconstruct the backdoor trigger instead of private trainers. Each class $j$ has its own trigger pattern $\epsilon_j$. For image $\mathbf{x}_i$ with label $i$, the backdoor image $\mathbf{x}_{i\rightarrow j}$ with label $j$ can be calculate as blended backdoor method:
\begin{equation}
    \mathbf{x}_{i\rightarrow j} = (1- \lambda_1 )\mathbf{x}_i + \lambda_1 \epsilon_j,
\end{equation}
where $\lambda_1$ is a hyperparameter. The training loss of the model can be calculate as:
\begin{equation}
    \mathcal{L}_{Trap}=(1-\lambda_2)\mathcal{L}_{CE}(\mathbf{y}, \hat{\mathbf{y}})+\lambda_2\mathcal{L}_{CE}(\mathbf{y}', \hat{\mathbf{y}}'),
\end{equation}
where $\mathbf{y}$ and $\hat{\mathbf{y}}$ are benign labels and predictions, $\mathbf{y}'$ and $\hat{\mathbf{y}}'$ are backdoor ones, and $\lambda_2$ is a hyperparameter.

\section{Back-propagation functions of some common non-linear activations.}
\label{appx:backward function}

\begin{figure}[!htbp]
    \centering
    \includegraphics[width=\linewidth]{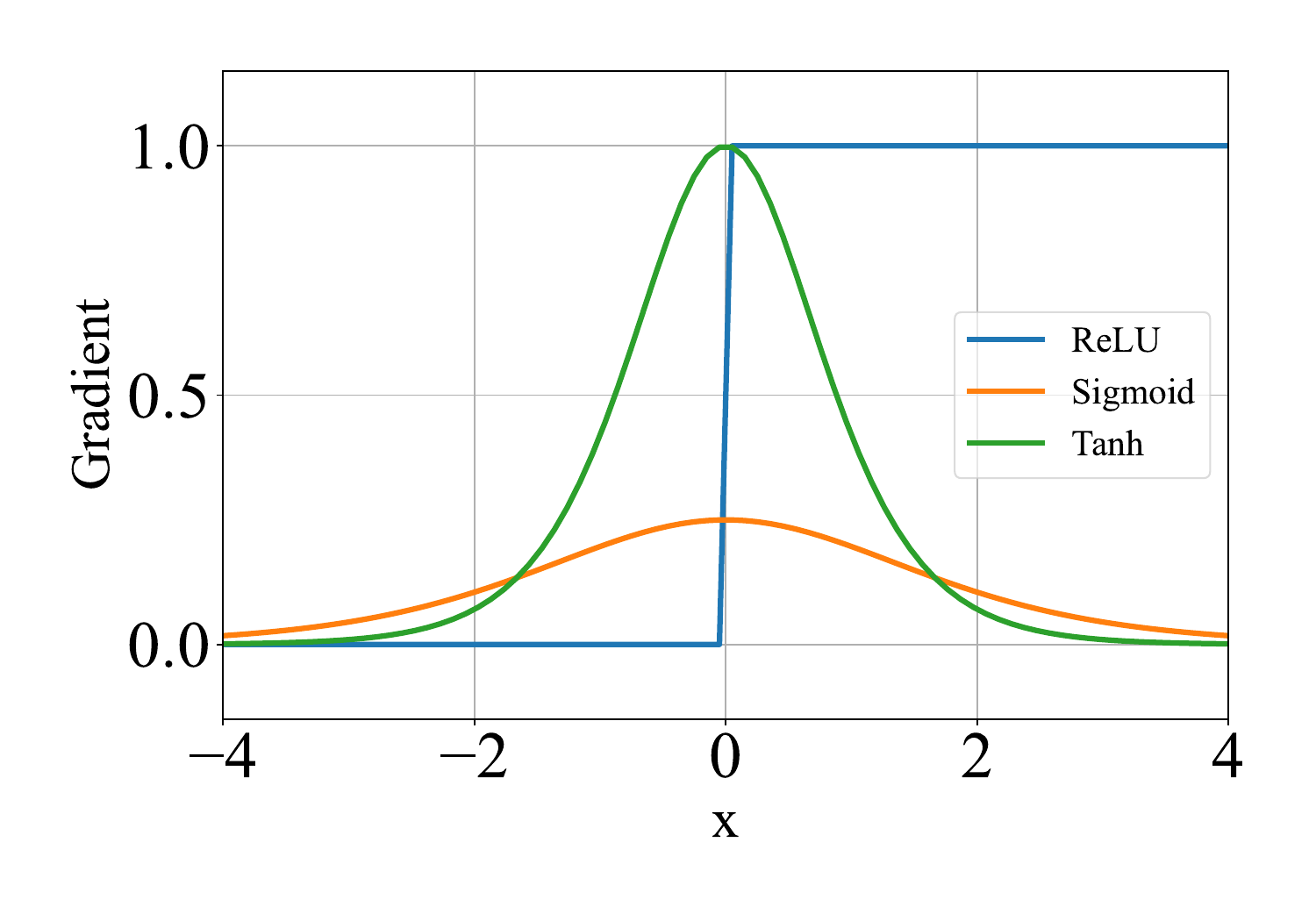}
    \caption{Back-propagation functions of some common non-linear activations.}
    \label{fig:backward function}
\end{figure}

Fig. \ref{fig:backward function} shows the back-propagation functions of some common non-linear activations, including $\mathrm{ReLU}$, $\mathrm{Sigmoid}$ and $\mathrm{Tanh}$. It shows that when the independent variable $x$ gets larger, the gradient of $\mathrm{ReLU}$ maintain at a very high level, and that of $\mathrm{Tanh}$ and $\mathrm{Sigmoid}$ gradually converging to $0$. 
$\mathrm{Tanh}$ converges to $0$ faster, meaning that it is more likely to cause the gradient vanishing problem. 

\section{Experiment Details}
\label{appx:exp detail}

\subsection{Hardware and Software Details}

For experiments in low-resolution scenarios, we conduct them on an Intel Xeon Platinum 8338C CPU with an NVIDIA RTX 4090 GPU. The training time of LoFt models is about 50 minutes. In high-resolution scenarios, we perform experiments on an Intel Xeon Gold 6330 CPU with an NVIDIA RTX A6000 GPU. The training time of LoFt models is about 150 minutes. Our code is available in the Supplementary Materials and will be open-sourced after publishing. 

\subsection{Classifier Training}
\label{appx:classifier training}

We train the classifier models based on prior research, incorporating both the low-resolution scenario \cite{gmi} and the high-resolution scenario \cite{ppa}.

Following previous researches \cite{gmi, ppa}, the origin weights in the low-resolution are provided by Face.evoLVe \cite{faceevolve}, and those in the high-resolution scenarios are the default weights of torchvision models \cite{torchvision}. To achieve high performance of classifiers, we pre-train the models on a large public dataset provided by \cite{msceleb1m} for $5$ epochs.

In low-resolution scenarios, we use the SGD optimizer \cite{sgd} with an initial learning rate of $0.01$ and momentum of $0.9$. The batch size is set to $128$. All data samples are cropped and resized to $64\times 64$ for IR-$152$ and $112\times 112$ for FaceNet-$112$, with a random horizontal flip applied with a probability of $50\%$. 

For high-resolution scenarios, the Adam optimizer \cite{adam} is employed with an initial learning rate of $0.001$ and $\beta = (0.9, 0.999)$. The batch size is $96$. All data samples are normalized with $\mu = 0.5$ and $\sigma = 0.5$, and resized to $224\times 224$. The training samples are augmented through random cropping within a scale range of $[0.85, 1.0]$ and a fixed aspect ratio of $1.0$. Crops are resized back to $224\times 224$, followed by random color jittering with brightness and contrast factors of $0.2$, and saturation and hue factors of $0.1$. A horizontal flip is applied with a probability of $50\%$.

In both scenarios, the models are trained for $100$ epochs, with the learning rate reduced by a factor of $0.1$ after epochs $75$ and $90$.

For the target classifier, to ensure a fair comparison of different defense algorithms, we meticulously tuned the hyperparameters of each algorithm to maintain nearly identical classification performance across models. 
The hyperparameters of previous defense methods are described in Appendix \ref{appx:defense summary}. For our defense, the adjustable hyperparameter is the dimension $r$ in the LoFt.
The specific hyperparameters used in the main paper are presented in Table \ref{tab:hyperparameters}. In high-resolution scenarios without pre-training, to reduce the impact of $\mathrm{Tanh}$’s vanishing gradient effect on training, we first trained for $50$ epochs without defense and then replaced the classification head with the LoFt classification head. 

For the evaluation models, we use FaceNet-$112$ and MaxViT in low- and high-resolution scenarios, respectively. The classification accuracy in the test dataset are shown in Table \ref{tab:eval test acc}.

\begin{table}[!ht]
    \setlength{\tabcolsep}{5pt}
    \normalsize
    \centering
    \caption{Hyperparameters for the experimental settings in the main paper. The models are trained on the FaceScrub dataset, with the IR-$152$ architecture used in the low-resolution scenario and ResNet-$152$ in the high-resolution scenario. The settings in the table are as follows: (A) Low-resolution scenario, (B) High-resolution scenario with high test accuracy, and (C) High-resolution scenario with low test accuracy. }
    \label{tab:hyperparameters}
    \begin{threeparttable} 
    \resizebox{\linewidth}{!}{
    \begin{tabular}{cccc}
        \toprule
\textbf{Settings} & \textbf{A} & \textbf{B} & \textbf{C}  \\ \midrule
\textbf{MID} & $0.1$ & $0.005$ &  $0.005$ 
\\
\textbf{BiDO} & $0.01, 0.1$ & $0.15, 1.5$ & $0.05, 0.5$
\\
\textbf{LS} & $-0.3$ & $-0.01$ & $-0.02$
\\
\textbf{TL} & $50\%$ & $70\%$ & $60\%$
\\
\textbf{RoLSS} & $2$ & $1$ & $1$ \\
\textbf{Trap} & $0.05, 0.3$ & $0.02, 0.2$ & $0.05, 0.2$ \\ 
\textbf{LoFt (ours)} & $30$ & $50$ & $32$
\\
\bottomrule
    \end{tabular}
}
    \end{threeparttable}
\end{table}
\begin{table}[!ht]
    \setlength{\tabcolsep}{5pt}
    \normalsize
    \centering
    \caption{The test classification accuracy of the evaluation models on different private datasets used in our experiments. FaceNet-$112$ is utilized for evaluation in the low-resolution scenario, while MaxViT is employed for the high-resolution scenario.}
    \label{tab:eval test acc}
    \begin{threeparttable} 
    \resizebox{.8\linewidth}{!}{
    \begin{tabular}{ccc}
        \toprule
\textbf{Test Acc} & \textbf{FaceScrub} & \textbf{CelebA}  \\ \midrule
\textbf{FaceNet-$112$} & $99.38\%$ & $95.88\%$
\\
\textbf{MaxViT} & $99.41\%$ & $97.23\%$
\\
\bottomrule
    \end{tabular}
}
    \end{threeparttable}
\end{table}

\subsection{Attacks}
\label{appx:attacks}

We perform various kinds of MI attacks, including GMI \cite{gmi}, KED \cite{ked}, Mirror \cite{mirror}, PPA \cite{ppa}, LOMMA \cite{lomma}, LOKT \cite{lokt},  PLG \cite{plg}, IF \cite{ifgmi} and PPDG \cite{ppdg}. Since the PPDG has three variants, we adopt the strongest one, \textit{i.e.}, PPDG-MMD. Following previous research \cite{gmi, plg}, we reconstruct $5$ images per class. Due to the high time costs for MIAs, we performed attacks on the first $100$ classes.

In our experimental setup, the prior dataset used by attackers is FFHQ, in line with previous researches \cite{ls,tl}. Attackers leverage this dataset to train GANs or surrogate classifiers \cite{lomma}. Specifically, GMI, KED, LOMMA, LOKT and PLG train customized GANs. This GANs are originally designed for low-resolution scenarios. To adapt to the high-resolution scenarios, we add two upsample blocks in the generator and two downsample blocks in the discriminator, following the settings in \cite{ppa}. Therefore, the generator can generate images with $256\times 256$ resolution, and the image will be resized to $224\times224$ before sent to the classifiers. Mirror, PPA, IF, and DDPG utilize a pre-trained FFHQ StyleGAN2-Ada model provided by \cite{stylegan2}. They mainly focus on high-resolution scenarios, with GANs at a resolution of $1024\times 1024$. The generated images will be cropped to $800\times 800$ and resized to $224\times 224$. To modify into low-resolution scenarios, we adopt the $256\times 256$ StyleGAN2-Ada, and the images will be cropped to $176\times 176$ and resized to $64\times 64$ for IR-$152$ or $112\times 112$ for FaceNet-$112$. For other attack settings and hyperparameters, we follow the official settings.


\section{Additional Experiment Results}
\label{appx:additional result}

In this section, we present some additional experimental results.

\paragraph{More attack methods.} We evaluate GMI \cite{gmi}, KED \cite{ked}, LOMMA \cite{lomma}, Mirror \cite{mirror}, PPA \cite{ppa} and PLG \cite{plg} in low-resolution scenarios against IR-$152$ and PLG in high-resolution scenarios against ResNet-$152$. The results are presented in Table \ref{tab:low ffhq facescrub gmi ked}, \ref{tab:low ffhq facescrub mirror ppa}, \ref{tab:low ffhq facescrub lomma}, \ref{tab:low ffhq facescrub plg}, \ref{tab:high ffhq facescrub plg im} and \ref{tab:high ffhq facescrub plg msceleb}. 
The results show that our defense is significantly more effective than previous methods in most scenarios.

\paragraph{Evaluation on the CelebA dataset.}

We evaluated the defense methods on the CelebA dataset under both low and high-resolution scenarios. The results, presented in Table 
\ref{tab:high ffhq celeba if resnet152} and \ref{tab:high ffhq celeba if}, indicate that our method consistently outperforms all other defense algorithms. This demonstrates the robustness of our approach across different datasets and resolutions.

\paragraph{Evaluation on transformer-based classifiers.}
In the previous experiments, we analyzed the impact of defense algorithms on convolution-based classifiers. To further explore this, we extended our experiments to include transformer-based models. The attack results in high-resolution scenarios are shown in Table \ref{tab:high ffhq facescrub if swin} and \ref{tab:high ffhq celeba if}. The results reveal that while the defense algorithm is generally less effective for transformer-based model structures, our method continues to achieve SOTA results.

\begin{table*}[!ht]
    \setlength{\tabcolsep}{5pt}
    \normalsize
    \centering
    \caption{GMI and KED attack results against IR-$152$ models trained on FaceScrub dataset in the low-resolution scenario.}
    \label{tab:low ffhq facescrub gmi ked}
    \begin{threeparttable} 
    \resizebox{\linewidth}{!}{
    \begin{tabular}{cccccccccc}
        \toprule
 \multirow{2}{*}{\textbf{Method}} 
 &  \multirow{2}{*}{\textbf{Test Acc}} 
 & \multicolumn{4}{c}{GMI} &\multicolumn{4}{c}{KED} \\
 \cmidrule(lr){3-6}\cmidrule(lr){7-10} & & 
 $\downarrow$ \textbf{Acc@1} & $\downarrow$ \textbf{Acc@5} & $\uparrow$ \textbf{$\delta_{eval}$} & $\uparrow$ \textbf{$\delta_{face}$} & $\downarrow$ \textbf{Acc@1} & $\downarrow$ \textbf{Acc@5} & $\uparrow$ \textbf{$\delta_{eval}$} & $\uparrow$ \textbf{$\delta_{face}$}  \\ \midrule
\textbf{NO Defense} & $98.2$ & $8.0_{\pm 8.6}$ & $11.6_{\pm 11.5}$ & $2466_{\pm 229}$ & $1.45_{\pm 0.18}$ & $37.8_{\pm 3.5}$ & $52.6_{\pm 4.8}$ & $2114_{\pm 546}$ & $1.04_{\pm 0.36}$ \\
\textbf{MID} & $97.0$ & $11.2_{\pm 7.2}$ & $20.6_{\pm 12.1}$ & $2507_{\pm 268}$ & $1.32_{\pm 0.17}$ & $60.0_{\pm 2.1}$ & $77.6_{\pm 2.7}$ & $1846_{\pm 454}$ & $0.81_{\pm 0.23}$ \\
\textbf{BiDO} & $95.2$ & $3.6_{\pm 4.1}$ & $12.6_{\pm 6.2}$ & $2390_{\pm 211}$ & $1.22_{\pm 0.14}$ & $20.2_{\pm 2.6}$ & $40.6_{\pm 2.8}$ & $2257_{\pm 399}$ & $0.98_{\pm 0.20}$ \\
\textbf{LS} & $97.3$ & $9.2_{\pm 5.8}$ & $20.6_{\pm 12.6}$ & $2482_{\pm 237}$ & $1.36_{\pm 0.15}$ & $26.8_{\pm 3.7}$ & $48.6_{\pm 1.9}$ & $2346_{\pm 572}$ & $1.12_{\pm 0.32}$ \\
\textbf{TL} & $95.4$ & $\mathbf{1.8}_{\pm 1.5}$ & $\mathbf{5.0}_{\pm 3.0}$ & $\mathbf{2694}_{\pm 270}$ & $1.46_{\pm 0.16}$ & $21.8_{\pm 1.8}$ & $33.8_{\pm 2.3}$ & $2280_{\pm 505}$ & $1.06_{\pm 0.28}$ \\
\textbf{RoLSS} & $97.4$ & $12.0_{\pm 6.6}$ & $24.2_{\pm 12.5}$ & $2433_{\pm 250}$ & $1.23_{\pm 0.15}$ & $32.2_{\pm 1.8}$ & $53.6_{\pm 3.3}$ & $2099_{\pm 420}$ & $0.95_{\pm 0.21}$ \\
\textbf{Trap} & $96.3$ & $4.6_{\pm 4.6}$ & $11.4_{\pm 10.4}$ & $2504_{\pm 269}$ & $1.25_{\pm 0.14}$ & $\mathbf{4.4}_{\pm 1.6}$ & $\mathbf{7.2}_{\pm 1.3}$ & $\mathbf{2745}_{\pm 415}$ & $\mathbf{1.45}_{\pm 0.29}$ \\
\textbf{LoFt (ours)} & $97.1$ & $3.4_{\pm 3.1}$ & $7.8_{\pm 6.9}$ & $2407_{\pm 222}$ & $\mathbf{1.50}_{\pm 0.15}$ & $9.6_{\pm 3.1}$ & $19.8_{\pm 2.9}$ & $2545_{\pm 443}$ & $1.32_{\pm 0.33}$ \\

\bottomrule
    \end{tabular}
    }
    \end{threeparttable}
\end{table*}
\begin{table*}[!ht]
    \setlength{\tabcolsep}{5pt}
    \normalsize
    \centering
    \caption{Mirror and PPA attack results against IR-$152$ models trained on FaceScrub dataset in the low-resolution scenario. }
    \label{tab:low ffhq facescrub mirror ppa}
    \begin{threeparttable} 
    \resizebox{\linewidth}{!}{
    \begin{tabular}{cccccccccc}
        \toprule
 \multirow{2}{*}{\textbf{Method}} 
 &  \multirow{2}{*}{\textbf{Test Acc}} 
 & \multicolumn{4}{c}{Mirror} &\multicolumn{4}{c}{PPA} \\
 \cmidrule(lr){3-6}\cmidrule(lr){7-10} & & 
 $\downarrow$ \textbf{Acc@1} & $\downarrow$ \textbf{Acc@5} & $\uparrow$ \textbf{$\delta_{eval}$} & $\uparrow$ \textbf{$\delta_{face}$} & $\downarrow$ \textbf{Acc@1} & $\downarrow$ \textbf{Acc@5} & $\uparrow$ \textbf{$\delta_{eval}$} & $\uparrow$ \textbf{$\delta_{face}$}  \\ \midrule

\textbf{NO Defense} & $98.2$ & $40.4_{\pm 5.5}$ & $57.4_{\pm 3.9}$ & $1911_{\pm 451}$ & $0.95_{\pm 0.26}$ & $91.2_{\pm 4.0}$ & $95.2_{\pm 2.7}$ & $1204_{\pm 335}$ & $0.63_{\pm 0.20}$ \\
\textbf{MID} & $97.0$ & $45.2_{\pm 4.0}$ & $59.2_{\pm 1.0}$ & $1919_{\pm 476}$ & $0.97_{\pm 0.23}$ & $90.4_{\pm 3.4}$ & $95.0_{\pm 2.2}$ & $1248_{\pm 284}$ & $0.65_{\pm 0.14}$ \\
\textbf{BiDO} & $95.2$ & $13.8_{\pm 0.7}$ & $27.8_{\pm 2.8}$ & $2271_{\pm 366}$ & $1.06_{\pm 0.18}$ & $53.6_{\pm 4.5}$ & $77.2_{\pm 2.2}$ & $1709_{\pm 367}$ & $0.78_{\pm 0.15}$ \\
\textbf{LS} & $97.3$ & $35.6_{\pm 2.9}$ & $58.4_{\pm 5.6}$ & $1938_{\pm 414}$ & $0.98_{\pm 0.27}$ & $91.6_{\pm 2.6}$ & $94.4_{\pm 2.7}$ & $1247_{\pm 362}$ & $0.66_{\pm 0.21}$ \\
\textbf{TL} & $95.4$ & $14.4_{\pm 3.1}$ & $29.4_{\pm 6.1}$ & $2231_{\pm 348}$ & $1.13_{\pm 0.19}$ & $57.0_{\pm 5.9}$ & $77.0_{\pm 8.0}$ & $1627_{\pm 427}$ & $0.78_{\pm 0.22}$ \\
\textbf{RoLSS} & $97.4$ & $19.8_{\pm 2.3}$ & $35.2_{\pm 1.7}$ & $2110_{\pm 438}$ & $1.09_{\pm 0.22}$ & $63.6_{\pm 5.1}$ & $82.6_{\pm 5.5}$ & $1567_{\pm 318}$ & $0.78_{\pm 0.16}$ \\
\textbf{Trap} & $96.3$ & $\mathbf{1.0}_{\pm 0.9}$ & $\mathbf{4.0}_{\pm 1.7}$ & $2498_{\pm 292}$ & $1.40_{\pm 0.19}$ & $42.2_{\pm 8.2}$ & $58.6_{\pm 7.1}$ & $1867_{\pm 501}$ & $0.95_{\pm 0.28}$ \\
\textbf{LoFt (ours)} & $97.1$ & $7.8_{\pm 1.8}$ & $14.0_{\pm 2.8}$ & $\mathbf{2543}_{\pm 394}$ & $\mathbf{1.41}_{\pm 0.29}$ & $\mathbf{22.2}_{\pm 6.0}$ & $\mathbf{37.2}_{\pm 8.2}$ & $\mathbf{2095}_{\pm 415}$ & $\mathbf{1.14}_{\pm 0.32}$ \\

\bottomrule
    \end{tabular}
    }
    \end{threeparttable}
\end{table*}
\begin{table*}[!ht]
    \setlength{\tabcolsep}{5pt}
    \normalsize
    \centering
    \caption{LOMMA attack results against IR-$152$ models trained on FaceScrub dataset in the low-resolution scenario.  Note that LOMMA is a plug-and-play technique that can seamlessly combine with existing generative model inversion attacks. In our experiments, we adhered to the official configurations, integrating LOMMA with both GMI and KED.}
    \label{tab:low ffhq facescrub lomma}
    \begin{threeparttable} 
    \resizebox{\linewidth}{!}{
    \begin{tabular}{cccccccccc}
        \toprule
 \multirow{2}{*}{\textbf{Method}} 
 &  \multirow{2}{*}{\textbf{Test Acc}} 
 & \multicolumn{4}{c}{LOMMA+GMI} &\multicolumn{4}{c}{LOMMA+KED} \\
 \cmidrule(lr){3-6}\cmidrule(lr){7-10} & & 
 $\downarrow$ \textbf{Acc@1} & $\downarrow$ \textbf{Acc@5} & $\uparrow$ \textbf{$\delta_{eval}$} & $\uparrow$ \textbf{$\delta_{face}$} & $\downarrow$ \textbf{Acc@1} & $\downarrow$ \textbf{Acc@5} & $\uparrow$ \textbf{$\delta_{eval}$} & $\uparrow$ \textbf{$\delta_{face}$}  \\ \midrule
\textbf{NO Defense} & $98.2$ & $48.0_{\pm 14.6}$ & $59.1_{\pm 11.1}$ & $2064_{\pm 342}$ & $1.05_{\pm 0.21}$ & $75.1_{\pm 0.8}$ & $87.2_{\pm 0.5}$ & $1641_{\pm 480}$ & $0.72_{\pm 0.27}$ \\
\textbf{MID} & $97.0$ & $56.7_{\pm 9.0}$ & $75.1_{\pm 7.2}$ & $1879_{\pm 303}$ & $0.83_{\pm 0.17}$ & $66.6_{\pm 1.9}$ & $86.6_{\pm 1.9}$ & $1708_{\pm 412}$ & $0.72_{\pm 0.20}$ \\
\textbf{BiDO} & $95.2$ & $29.8_{\pm 13.4}$ & $48.6_{\pm 17.7}$ & $2254_{\pm 288}$ & $0.92_{\pm 0.15}$ & $64.6_{\pm 1.4}$ & $83.6_{\pm 3.3}$ & $1782_{\pm 365}$ & $0.72_{\pm 0.17}$ \\
\textbf{LS} & $97.3$ & $50.8_{\pm 17.6}$ & $64.9_{\pm 16.9}$ & $2096_{\pm 370}$ & $0.99_{\pm 0.21}$ & $78.6_{\pm 1.0}$ & $90.4_{\pm 0.8}$ & $1526_{\pm 434}$ & $0.70_{\pm 0.22}$ \\
\textbf{TL} & $95.4$ & $\mathbf{14.9}_{\pm 7.9}$ & $28.5_{\pm 11.1}$ & $2387_{\pm 319}$ & $1.13_{\pm 0.21}$ & $38.8_{\pm 1.6}$ & $58.4_{\pm 1.0}$ & $2031_{\pm 559}$ & $0.91_{\pm 0.30}$ \\
\textbf{RoLSS} & $97.4$ & $36.4_{\pm 9.0}$ & $54.8_{\pm 9.0}$ & $2095_{\pm 258}$ & $0.94_{\pm 0.19}$ & $49.6_{\pm 1.5}$ & $71.0_{\pm 3.0}$ & $1902_{\pm 439}$ & $0.84_{\pm 0.26}$ \\
\textbf{Trap} & $96.3$ & $38.0_{\pm 6.9}$ & $62.6_{\pm 9.4}$ & $1988_{\pm 262}$ & $0.81_{\pm 0.12}$ & $42.6_{\pm 2.7}$ & $60.4_{\pm 2.4}$ & $1954_{\pm 538}$ & $0.80_{\pm 0.24}$ \\
\textbf{LoFt (ours)} & $97.1$ & $15.2_{\pm 8.6}$ & $\mathbf{24.6}_{\pm 12.5}$ & $\mathbf{2460}_{\pm 307}$ & $\mathbf{1.27}_{\pm 0.23}$ & $\mathbf{35.0}_{\pm 2.4}$ & $\mathbf{56.6}_{\pm 1.6}$ & $\mathbf{2089}_{\pm 543}$ & $\mathbf{0.98}_{\pm 0.38}$ \\

\bottomrule
    \end{tabular}
    }
    \end{threeparttable}
\end{table*}
\begin{table}[!ht]
    \setlength{\tabcolsep}{5pt}
    \normalsize
    \centering
    \caption{PLG attack results against IR-$152$ models trained on FaceScrub dataset in the low-resolution scenario.}
    \label{tab:low ffhq facescrub plg}
    \begin{threeparttable} 
    \resizebox{\linewidth}{!}{
    \begin{tabular}{cccccccccc}
        \toprule
 \multirow{2}{*}{\textbf{Method}} 
 &  \multirow{2}{*}{\textbf{Test Acc}} 
 & \multicolumn{4}{c}{PLG}\\
 \cmidrule(lr){3-6} & & 
 $\downarrow$ \textbf{Acc@1} & $\downarrow$ \textbf{Acc@5} & $\uparrow$ \textbf{$\delta_{eval}$} & $\uparrow$ \textbf{$\delta_{face}$}  \\ \midrule
\textbf{NO Defense} & $98.2$ & $100.0_{\pm 0.0}$ & $100.0_{\pm 0.0}$ & $908_{\pm 176}$ & $0.47_{\pm 0.09}$ \\
\textbf{MID} & $97.0$ & $99.2_{\pm 1.0}$ & $99.6_{\pm 0.8}$ & $1063_{\pm 211}$ & $0.54_{\pm 0.11}$ \\
\textbf{BiDO} & $95.2$ & $82.2_{\pm 2.0}$ & $92.6_{\pm 2.9}$ & $1533_{\pm 292}$ & $0.63_{\pm 0.14}$ \\
\textbf{LS} & $97.3$ & $99.8_{\pm 0.4}$ & $100.0_{\pm 0.0}$ & $840_{\pm 205}$ & $0.46_{\pm 0.10}$ \\
\textbf{TL} & $95.4$ & $93.6_{\pm 1.6}$ & $97.8_{\pm 1.0}$ & $1202_{\pm 311}$ & $0.54_{\pm 0.13}$ \\
\textbf{RoLSS} & $97.4$ & $89.4_{\pm 2.7}$ & $96.6_{\pm 1.0}$ & $1398_{\pm 268}$ & $0.63_{\pm 0.11}$ \\
\textbf{Trap} & $96.3$ & $80.2_{\pm 2.8}$ & $91.2_{\pm 1.8}$ & $1559_{\pm 304}$ & $0.64_{\pm 0.12}$ \\
\textbf{LoFt (ours)} & $97.1$ & $\mathbf{46.8}_{\pm 8.3}$ & $\mathbf{63.0}_{\pm 8.4}$ & $\mathbf{2009}_{\pm 403}$ & $\mathbf{0.96}_{\pm 0.28}$ \\
\bottomrule
    \end{tabular}
}
    \end{threeparttable}
\end{table}
\begin{table}[!h]
    \setlength{\tabcolsep}{5pt}
    \normalsize
    \centering
    \caption{IF attack results against Swin-v2 models trained on FaceScrub dataset in the high-resolution scenario.}
    \label{tab:high ffhq facescrub if swin}
    \begin{threeparttable} 
    \resizebox{\linewidth}{!}{
    \begin{tabular}{cccccccccc}
        \toprule
 \multirow{2}{*}{\textbf{Method}} 
 &  \multirow{2}{*}{\textbf{Test Acc}} 
 & \multicolumn{4}{c}{IF}\\
 \cmidrule(lr){3-6} & & 
 $\downarrow$ \textbf{Acc@1} & $\downarrow$ \textbf{Acc@5} & $\uparrow$ \textbf{$\delta_{eval}$} & $\uparrow$ \textbf{$\delta_{face}$}  \\ \midrule

\textbf{NO Defense} & $97.8$ & $94.0_{\pm 0.6}$ & $97.6_{\pm 0.8}$ & $354_{\pm 63}$ & $0.66_{\pm 0.14}$ \\
\textbf{MID} & $97.3$ & $77.6_{\pm 3.8}$ & $90.2_{\pm 2.7}$ & $417_{\pm 68}$ & $0.82_{\pm 0.14}$ \\
\textbf{BiDO} & $97.9$ & $95.0_{\pm 1.3}$ & $97.4_{\pm 0.8}$ & $345_{\pm 67}$ & $0.64_{\pm 0.15}$ \\
\textbf{LS} & $97.6$ & $94.6_{\pm 0.5}$ & $99.6_{\pm 0.8}$ & $374_{\pm 55}$ & $0.68_{\pm 0.13}$ \\
\textbf{TL} & $98.2$ & $94.2_{\pm 2.8}$ & $98.2_{\pm 0.7}$ & $342_{\pm 62}$ & $0.63_{\pm 0.14}$ \\
\textbf{RoLSS} & $97.0$ & $95.2_{\pm 1.9}$ & $98.4_{\pm 0.8}$ & $349_{\pm 58}$ & $0.64_{\pm 0.13}$ \\
\textbf{Trap} & $95.1$ & $86.0_{\pm 0.9}$ & $90.8_{\pm 1.6}$ & $374_{\pm 107}$ & $0.70_{\pm 0.23}$ \\
\textbf{LoFt (ours)} & $97.4$ & $\mathbf{64.4}_{\pm 3.9}$ & $\mathbf{84.2}_{\pm 1.6}$ & $\mathbf{434}_{\pm 68}$ & $\mathbf{0.85}_{\pm 0.16}$ \\

\bottomrule
    \end{tabular}
}
    \end{threeparttable}
\end{table}
\begin{table}[!ht]
    \setlength{\tabcolsep}{5pt}
    \normalsize
    \centering
    \caption{IF attack results against ResNet-$152$ models trained on CelebA dataset in the high-resolution scenario.}
    \label{tab:high ffhq celeba if resnet152}
    \begin{threeparttable} 
    \resizebox{\linewidth}{!}{
    \begin{tabular}{cccccccccc}
        \toprule
 \multirow{2}{*}{\textbf{Method}} 
 &  \multirow{2}{*}{\textbf{Test Acc}} 
 & \multicolumn{4}{c}{IF}\\
 \cmidrule(lr){3-6} & & 
 $\downarrow$ \textbf{Acc@1} & $\downarrow$ \textbf{Acc@5} & $\uparrow$ \textbf{$\delta_{eval}$} & $\uparrow$ \textbf{$\delta_{face}$}  \\ \midrule

\textbf{NO Defense} & $96.3$ & $99.8_{\pm 0.4}$ & $100.0_{\pm 0.0}$ & $277_{\pm 55}$ & $0.46_{\pm 0.12}$ \\
\textbf{MID} & $94.3$ & $98.6_{\pm 1.0}$ & $99.8_{\pm 0.4}$ & $266_{\pm 61}$ & $0.44_{\pm 0.13}$ \\
\textbf{BiDO} & $94.7$ & $97.8_{\pm 1.2}$ & $99.6_{\pm 0.5}$ & $293_{\pm 57}$ & $0.48_{\pm 0.11}$ \\
\textbf{LS} & $93.6$ & $92.6_{\pm 2.1}$ & $98.8_{\pm 1.2}$ & $342_{\pm 76}$ & $0.61_{\pm 0.18}$ \\
\textbf{TL} & $94.1$ & $100.0_{\pm 0.0}$ & $100.0_{\pm 0.0}$ & $229_{\pm 44}$ & $0.37_{\pm 0.09}$ \\
\textbf{RoLSS} & $93.8$ & $82.8_{\pm 1.2}$ & $96.0_{\pm 1.4}$ & $359_{\pm 59}$ & $0.64_{\pm 0.14}$ \\
\textbf{Trap} & $92.1$ & $70.2_{\pm 2.3}$ & $87.2_{\pm 1.2}$ & $398_{\pm 87}$ & $0.71_{\pm 0.20}$ \\
\textbf{LoFt (ours)} & $94.4$ & $\mathbf{62.6}_{\pm 3.3}$ & $\mathbf{79.6}_{\pm 1.9}$ & $\mathbf{414}_{\pm 85}$ & $\mathbf{0.79}_{\pm 0.20}$ \\

\bottomrule
    \end{tabular}
}
    \end{threeparttable}
\end{table}
\begin{table}[!ht]
    \setlength{\tabcolsep}{5pt}
    \normalsize
    \centering
    \caption{PLG attack results against ResNet-$152$ models with low accuracy on FaceScrub dataset in the high-resolution scenario.}
    \label{tab:high ffhq facescrub plg im}
    \begin{threeparttable} 
    \resizebox{\linewidth}{!}{
    \begin{tabular}{cccccccccc}
        \toprule
 \multirow{2}{*}{\textbf{Method}} 
 &  \multirow{2}{*}{\textbf{Test Acc}} 
 & \multicolumn{4}{c}{PLG}\\
 \cmidrule(lr){3-6} & & 
 $\downarrow$ \textbf{Acc@1} & $\downarrow$ \textbf{Acc@5} & $\uparrow$ \textbf{$\delta_{eval}$} & $\uparrow$ \textbf{$\delta_{face}$}  \\ \midrule
\textbf{NO Defense} & $92.2$ & $99.2_{\pm 0.7}$ & $100.0_{\pm 0.0}$ & $312_{\pm 49}$ & $0.55_{\pm 0.11}$ \\
\textbf{MID} & $88.2$ & $99.6_{\pm 0.8}$ & $100.0_{\pm 0.0}$ & $287_{\pm 38}$ & $0.53_{\pm 0.10}$ \\
\textbf{BiDO} & $88.6$ & $98.4_{\pm 0.8}$ & $99.8_{\pm 0.4}$ & $340_{\pm 58}$ & $0.58_{\pm 0.12}$ \\
\textbf{LS} & $88.8$ & $64.6_{\pm 4.3}$ & $73.2_{\pm 5.3}$ & $448_{\pm 83}$ & $0.92_{\pm 0.24}$ \\
\textbf{TL} & $88.4$ & $98.6_{\pm 0.8}$ & $99.8_{\pm 0.4}$ & $336_{\pm 46}$ & $0.59_{\pm 0.10}$ \\
\textbf{RoLSS} & $88.7$ & $98.8_{\pm 0.7}$ & $100.0_{\pm 0.0}$ & $323_{\pm 46}$ & $0.59_{\pm 0.11}$ \\
\textbf{Trap} & $85.7$ & $92.4_{\pm 2.9}$ & $98.0_{\pm 0.9}$ & $370_{\pm 52}$ & $0.69_{\pm 0.12}$ \\
\textbf{LoFt (ours)} & $89.4$ & $\mathbf{47.2}_{\pm 7.5}$ & $\mathbf{68.8}_{\pm 5.3}$ & $\mathbf{498}_{\pm 64}$ & $\mathbf{1.00}_{\pm 0.16}$ \\
\bottomrule
    \end{tabular}
}
    \end{threeparttable}
\end{table}

\begin{table}[!ht]
    \setlength{\tabcolsep}{5pt}
    \normalsize
    \centering
    \caption{PLG attack results against ResNet-$152$ models with high accuracy on FaceScrub dataset in the high-resolution scenario. }
    \label{tab:high ffhq facescrub plg msceleb}
    \begin{threeparttable} 
    \resizebox{\linewidth}{!}{
    \begin{tabular}{cccccccccc}
        \toprule
 \multirow{2}{*}{\textbf{Method}} 
 &  \multirow{2}{*}{\textbf{Test Acc}} 
 & \multicolumn{4}{c}{PLG}\\
 \cmidrule(lr){3-6} & & 
 $\downarrow$ \textbf{Acc@1} & $\downarrow$ \textbf{Acc@5} & $\uparrow$ \textbf{$\delta_{eval}$} & $\uparrow$ \textbf{$\delta_{face}$}  \\ \midrule
\textbf{NO Defense} & $98.5$ & $99.2_{\pm 0.7}$ & $100.0_{\pm 0.0}$ & $312_{\pm 49}$ & $0.55_{\pm 0.11}$ \\
\textbf{MID} & $96.8$ & $99.6_{\pm 0.8}$ & $100.0_{\pm 0.0}$ & $287_{\pm 38}$ & $0.53_{\pm 0.10}$ \\
\textbf{BiDO} & $96.3$ & $47.0_{\pm 4.6}$ & $67.2_{\pm 4.3}$ & $500_{\pm 66}$ & $1.01_{\pm 0.17}$ \\
\textbf{LS} & $96.4$ & $64.6_{\pm 4.3}$ & $73.2_{\pm 5.3}$ & $448_{\pm 83}$ & $0.92_{\pm 0.24}$ \\
\textbf{TL} & $96.6$ & $98.6_{\pm 0.8}$ & $99.8_{\pm 0.4}$ & $336_{\pm 46}$ & $0.59_{\pm 0.10}$ \\
\textbf{RoLSS} & $96.0$ & $98.8_{\pm 0.7}$ & $100.0_{\pm 0.0}$ & $323_{\pm 46}$ & $0.59_{\pm 0.11}$ \\
\textbf{Trap} & $96.7$ & $82.4_{\pm 4.2}$ & $95.0_{\pm 1.9}$ & $401_{\pm 71}$ & $0.75_{\pm 0.16}$ \\
\textbf{LoFt (ours)} & $96.7$ & $\mathbf{12.0}_{\pm 4.2}$ & $\mathbf{25.2}_{\pm 5.0}$ & $\mathbf{607}_{\pm 71}$ & $\mathbf{1.30}_{\pm 0.19}$ \\
\bottomrule
    \end{tabular}
}
    \end{threeparttable}
\end{table}
\begin{table}[!h]
    \setlength{\tabcolsep}{5pt}
    \normalsize
    \centering
    \caption{IF attack results against ViT-B/16 models trained on CelebA dataset in the high-resolution scenario. }
    \label{tab:high ffhq celeba if}
    \begin{threeparttable} 
    \resizebox{\linewidth}{!}{
    \begin{tabular}{cccccccccc}
        \toprule
 \multirow{2}{*}{\textbf{Method}} 
 &  \multirow{2}{*}{\textbf{Test Acc}} 
 & \multicolumn{4}{c}{IF}\\
 \cmidrule(lr){3-6} & & 
 $\downarrow$ \textbf{Acc@1} & $\downarrow$ \textbf{Acc@5} & $\uparrow$ \textbf{$\delta_{eval}$} & $\uparrow$ \textbf{$\delta_{face}$}  \\ \midrule
\textbf{NO Defense} & $95.4$ & $84.6_{\pm 2.7}$ & $94.2_{\pm 1.5}$ & $363_{\pm 83}$ & $0.65_{\pm 0.18}$ \\
\textbf{MID} & $92.6$ & $93.8_{\pm 2.6}$ & $97.0_{\pm 1.5}$ & $321_{\pm 72}$ & $0.56_{\pm 0.15}$ \\
\textbf{BiDO} & $91.4$ & $79.4_{\pm 2.5}$ & $89.2_{\pm 2.6}$ & $382_{\pm 83}$ & $0.69_{\pm 0.18}$ \\
\textbf{LS} & $92.2$ & $88.6_{\pm 2.7}$ & $97.0_{\pm 1.1}$ & $366_{\pm 59}$ & $0.62_{\pm 0.14}$ \\
\textbf{TL} & $95.6$ & $94.4_{\pm 2.2}$ & $97.4_{\pm 1.7}$ & $326_{\pm 62}$ & $0.56_{\pm 0.13}$ \\
\textbf{RoLSS} & $90.4$ & $75.8_{\pm 1.9}$ & $89.6_{\pm 0.8}$ & $389_{\pm 83}$ & $0.71_{\pm 0.18}$ \\
\textbf{Trap} & $91.8$ & $79.2_{\pm 2.5}$ & $\mathbf{86.4}_{\pm 2.7}$ & $\mathbf{403}_{\pm 129}$ & $0.70_{\pm 0.28}$ \\
\textbf{LoFt (ours)} & $91.9$ & $\mathbf{73.0}_{\pm 1.4}$ & $\mathbf{86.4}_{\pm 1.7}$ & $\mathbf{403}_{\pm 90}$ & $\mathbf{0.76}_{\pm 0.22}$ \\

\bottomrule
    \end{tabular}
}
    \end{threeparttable}
\end{table}

\section{Model Robustness in Other Aspects}
\label{appx:other robustness}

\subsection{Knowledge Extraction Score of MIAs}
\label{appx:kes}

Following the approach of \cite{ls}, we compute the Knowledge Extraction Score (KES), a metric designed to quantify the discriminative information extracted about distinct classes from MIAs. Specifically, we train a surrogate ResNet-$152$ \cite{resnet} classifier on the synthetic data generated by attacks and evaluate its classification accuracy on the target model’s original training set. The intuition behind KES is that a more successful inversion attack will enable the surrogate model to better differentiate between classes.

In our experiments, we employ synthetic attack data generated by IF attacks against a ResNet-$152$ \cite{resnet} model trained on the FaceScrub dataset. The results of these experiments are shown in Table \ref{tab:kes}. 
The results shows that the knowledge extraction score of IF attacks against classifiers with our defense method achieve the lowest results, which indicates that MIAs can hardly extract private information from models with our defense methods.

\begin{table}[!h]
    \setlength{\tabcolsep}{5pt}
    \normalsize
    \centering
    \caption{Knowledge extraction scores in high-resolution scenarios.}
    \label{tab:kes}
    \vspace{5pt}
    \begin{threeparttable} 
    \resizebox{\linewidth}{!}{
    \begin{tabular}{ccccccccccc}
        \toprule
        
 \multirow{2}{*}{\textbf{Method}}
 & \multicolumn{2}{c}{Pre-trained} &\multicolumn{2}{c}{Not Pre-trained} \\
 \cmidrule(lr){2-3}\cmidrule(lr){4-5} & 
  \textbf{Test Acc} & \textbf{$\downarrow$KES} & \textbf{Test Acc} & \textbf{$\downarrow$KES} \\\midrule

\textbf{NO Defense} & $98.5$ & $91.8$ & $92.2$ & $72.7$ \\
\textbf{MID} & $96.8$ & $90.2$ & $88.2$ & $83.6$ \\
\textbf{BiDO} & $96.3$ & $91.6$ & $88.6$ & $72.9$ \\
\textbf{LS} & $96.4$ & $90.9$ & $88.8$ & $68.4$ \\
\textbf{TL} & $96.6$ & $96.0$ & $88.4$ & $60.0$ \\
\textbf{RoLSS} & $96.0$ & $81.1$ & $88.7$ & $60.2$ \\
\textbf{Trap} & $96.7$ & $88.4$ & $85.7$ & $74.9$ \\
\textbf{LoFt (ours)} & $96.7$ & $\mathbf{64.9}$ & $89.4$ & $\mathbf{35.7}$ \\

\bottomrule
    \end{tabular}
    }
    \end{threeparttable}
\end{table}

\subsection{Adversarial Robustness}

In this section, we investigate if training with different defense methods has an impact on a model's robustness against adversarial attacks. Following \cite{ls}, we apply the following attacks to test model robustness:
\begin{itemize}
    \item \textbf{Fast Gradient Sign Method (FGSM) } \cite{fgsm}: One-step white-box attack. $\epsilon=2/255$.
    
    \item \textbf{Projected Gradient Descent (PGD) } \cite{pgd}: Multi-step white-box attack. 
        $\epsilon=2/255$, step size $=2/255$, steps $=10$, random start $=$ True.
    \item \textbf{Basic Iterative Method (BIM) } \cite{bim}: Multi-step white-box attack. $\epsilon=2/255$, step size $=2/255$, steps $=10$.
    \item \textbf{One Pixel Attack } \cite{onepixel}: Multi-step black-box attack. Pixels $=1$, steps $=10$, population size $=10$.
\end{itemize}

Adversarial attacks are conducted on test samples that are excluded from the training data. These attacks are evaluated in both targeted and untargeted settings. In the untargeted scenario, an attack is considered successful if the predicted label differs from the ground truth label. For targeted attacks, the target label is set to the original label plus one, and the attack is successful if the model predicts this target label. In both cases, we measure the attack success rate (ASR), where a lower ASR indicates greater robustness of the model to adversarial perturbations. The results in Table \ref{tab:adversarial} demonstrate that training a model with our defense method can make a model more robust to adversarial examples, especially in the targeted scenario.

\begin{table*}
    \setlength{\tabcolsep}{5pt}
    \normalsize
    \centering
    \caption{Adversarial attack results against ResNet-$152$ models trained on FaceScrub dataset in the high-resolution scenario.}
    \label{tab:adversarial}
    \begin{threeparttable} 
    \resizebox{\linewidth}{!}{
    \begin{tabular}{ccccccccccc}
        \toprule
 \multirow{2}{*}{\textbf{Pre-trained}} 
 & \multirow{2}{*}{\textbf{Method}} 
 &  \multirow{2}{*}{\textbf{Test Acc}} 
 & \multicolumn{4}{c}{Untargeted Attacks} &\multicolumn{4}{c}{Targeted Attacks} \\
 \cmidrule(lr){4-7}\cmidrule(lr){8-11} & & &
  $\downarrow$ \textbf{FGSM} & $\downarrow$ \textbf{PGD} & $\downarrow$ \textbf{BIM} & $\downarrow$ \textbf{OnePixel} & $\downarrow$ \textbf{FGSM} & $\downarrow$ \textbf{PGD} & $\downarrow$ \textbf{BIM} & $\downarrow$ \textbf{OnePixel} \\\midrule

\multirow{7}{*}{$\checkmark$}  & \textbf{NO Defense} & $98.5$ & $28.1$ & $65.8$ & $57.2$ & $\mathbf{0.0}$ & $75.5$ & $99.9$ & $99.8$ & $\mathbf{0.0}$ \\
 & \textbf{MID} & $96.8$ & $19.1$ & $50.1$ & $49.4$ & $\mathbf{0.0}$ & $79.8$ & $98.2$ & $98.4$ & $1.0$ \\
 & \textbf{BiDO} & $96.3$ & $22.6$ & $62.9$ & $57.1$ & $\mathbf{0.0}$ & $81.3$ & $99.9$ & $99.9$ & $\mathbf{0.0}$ \\
 & \textbf{LS} & $96.4$ & $20.1$ & $45.3$ & $38.6$ & $\mathbf{0.0}$ & $91.0$ & $99.6$ & $99.6$ & $1.4$ \\
 & \textbf{TL} & $96.6$ & $18.4$ & $49.7$ & $43.7$ & $\mathbf{0.0}$ & $76.9$ & $98.8$ & $98.7$ & $1.2$ \\
 & \textbf{RoLSS} & $96.0$ & $23.1$ & $56.8$ & $51.0$ & $\mathbf{0.0}$ & $90.2$ & $100.0$ & $100.0$ & $\mathbf{0.0}$ \\
 & \textbf{Trap} & $96.7$ & $6.9$ & $50.9$ & $55.7$ & $\mathbf{0.0}$ & $84.2$ & $\mathbf{80.4}$ & $\mathbf{86.5}$ & $2.6$ \\
 & \textbf{LoFt (ours)} & $96.7$ & $\mathbf{0.7}$ & $\mathbf{21.1}$ & $\mathbf{19.9}$ & $\mathbf{0.0}$ & $\mathbf{34.2}$ & $96.6$ & $96.8$ & $\mathbf{0.0}$ \\

\midrule

 \multirow{6}{*}{$\times$} & \textbf{NO Defense} & $92.2$ & $89.5$ & $100.0$ & $100.0$ & $11.5$ & $38.0$ & $79.0$ & $80.2$ & $0.5$ \\
 & \textbf{MID} & $88.2$ & $82.9$ & $99.8$ & $99.8$ & $12.7$ & $4.1$ & $48.8$ & $42.7$ & $\mathbf{0.0}$ \\
 & \textbf{BiDO} & $88.6$ & $89.5$ & $100.0$ & $100.0$ & $\mathbf{9.5}$ & $24.4$ & $62.7$ & $56.8$ & $0.2$ \\
 & \textbf{LS} & $88.8$ & $\mathbf{19.8}$ & $\mathbf{20.5}$ & $\mathbf{20.5}$ & $12.0$ & $26.1$ & $58.0$ & $55.1$ & $\mathbf{0.0}$ \\
 & \textbf{TL} & $88.4$ & $90.0$ & $100.0$ & $100.0$ & $12.9$ & $24.4$ & $84.6$ & $88.8$ & $0.2$ \\
 & \textbf{RoLSS} & $88.7$ & $97.6$ & $100.0$ & $100.0$ & $17.6$ & $24.4$ & $71.7$ & $71.7$ & $0.2$ \\
 & \textbf{Trap} & $85.7$ & $96.1$ & $99.5$ & $99.5$ & $32.0$ & $31.2$ & $79.5$ & $78.5$ & $0.2$ \\
 & \textbf{LoFt (ours)} & $89.4$ & $81.7$ & $99.5$ & $99.8$ & $9.8$ & $\mathbf{2.0}$ & $\mathbf{19.8}$ & $\mathbf{21.2}$ & $\mathbf{0.0}$ \\

\bottomrule
    \end{tabular}
    }
    \end{threeparttable}
\end{table*}

\subsection{Backdoor Robustness}

In addition to adversarial robustness, we also investigate of training with our proposed LoFt has an impact on backdoor attacks.  Due to the relatively small number of samples per class in FaceScrub and CelebA, it is not able to stably train the models \cite{ls}. Therefore, we 
train ResNet-$152$ models on a poisoned ImageNette \cite{imagenette} dataset, which is a subset of ImageNet \cite{imagenet} classes. Following the settings of \cite{ls}, we evaluate the backdoor robustness on the folowing attack methods:
\begin{itemize}
    \item \textbf{BadNets} \cite{badnets}: We add a $9\times 9$ checkerboard pattern to the lower right corner of reach image. In total, $1\%$ of all images are poisioned and labeled as class $0$.
    \item \textbf{Blended} \cite{blended}: We interpolate each poisoned image with a fixed Gaussian noise pattern. The blend ratio is set to 0.1. In total, $1\%$ of all images are poisoned and labeled as class 0.
\end{itemize}

For evaluation, we computed the classification accuracy on the test splits. To calculate the attack success rate, all test images are added with triggers. An poisoned image is viewed success if it is classified as the class $0$. Lower attack success rate means the more robust the model is to the backdoor attack.

\begin{table}[!ht]
    \setlength{\tabcolsep}{5pt}
    \normalsize
    \centering
    \caption{Backdoor attack results against ResNet-$152$ models trained on the ImageNette dataset.}
    \label{tab:backdoor}
    \vspace{5pt}
    \begin{threeparttable} 
    \resizebox{\linewidth}{!}{
    \begin{tabular}{cccc}
        \toprule
\textbf{Trigger} & \textbf{Defense} & $\uparrow$\textbf{Clean Accuracy} & $\downarrow$ \textbf{Attack Success Rate} \\
\midrule
 \multirow{2}{*}{\textbf{Clean}}  & \textbf{NO defense} & 83.1\% & - \\
 & \textbf{LoFt} & 80.5\% & - \\
\midrule
 \multirow{2}{*}{\textbf{BadNets}}  & \textbf{NO defense} & 79.7\% & 89.9\% \\
 & \textbf{LoFt} & 82.4\% & \textbf{88.6\%} \\
\midrule
 \multirow{2}{*}{\textbf{Blended}} & \textbf{NO defense} & 85.6\% & 87.4\% \\
 & \textbf{LoFt} & 77.9\% & \textbf{43.9\%} \\
\bottomrule
    \end{tabular}
    }
    \end{threeparttable}
\end{table}

\section{The Attack Results of Different Dimensions in High-Resolution Scenarios.}
\label{appx:rank result high}

Table \ref{tab:rank result high} presents the IF attack results on models with varying compressed dimensions. These models are trained on the FaceScrub dataset in the high-resolution scenario. The results demonstrate that a low dimension is sufficient to achieve high model performance, while increasing the dimension leads to a higher risk of privacy leakage.

\begin{table}[!ht]
    \setlength{\tabcolsep}{5pt}
    \normalsize
    \centering
    \caption{IF attack results against ResNet-$152$ models with different dimensions in the classifier head in high-resolution scenarios.}
    \label{tab:rank result high}
    \begin{threeparttable} 
    \resizebox{\linewidth}{!}{
    \begin{tabular}{cccccccccc}
        \toprule
{\textbf{Dimension}} 
 &  {\textbf{Test Acc}} &
 $\downarrow$ \textbf{Acc@1} & $\downarrow$ \textbf{Acc@5} & $\uparrow$ \textbf{$\delta_{eval}$} & $\uparrow$ \textbf{$\delta_{face}$}  \\ \midrule
\textbf{20} & $92.8$ & $18.6_{\pm 2.0}$ & $33.2_{\pm 2.1}$ & $581_{\pm 104}$ & $1.20_{\pm 0.24}$ \\
\textbf{35} & $95.4$ & $39.2_{\pm 6.7}$ & $60.6_{\pm 7.0}$ & $498_{\pm 87}$ & $1.01_{\pm 0.20}$ \\
\textbf{40} & $96.1$ & $51.2_{\pm 4.1}$ & $69.4_{\pm 3.5}$ & $480_{\pm 94}$ & $0.97_{\pm 0.23}$ \\
\textbf{50} & $96.6$ & $64.8_{\pm 4.5}$ & $83.2_{\pm 2.3}$ & $431_{\pm 74}$ & $0.86_{\pm 0.19}$ \\
\textbf{75} & $97.5$ & $80.4_{\pm 3.5}$ & $91.4_{\pm 2.2}$ & $392_{\pm 65}$ & $0.76_{\pm 0.15}$ \\
\textbf{100} & $97.7$ & $93.0_{\pm 2.8}$ & $97.6_{\pm 1.5}$ & $349_{\pm 53}$ & $0.66_{\pm 0.12}$ \\
\textbf{150} & $98.6$ & $97.8_{\pm 1.7}$ & $99.4_{\pm 0.5}$ & $321_{\pm 46}$ & $0.60_{\pm 0.10}$ \\
\textbf{200} & $98.7$ & $98.8_{\pm 0.7}$ & $100.0_{\pm 0.0}$ & $308_{\pm 41}$ & $0.58_{\pm 0.10}$ \\
\textbf{300} & $98.9$ & $98.2_{\pm 1.2}$ & $99.2_{\pm 0.7}$ & $304_{\pm 44}$ & $0.56_{\pm 0.11}$ \\
\textbf{500} & $98.9$ & $99.0_{\pm 0.6}$ & $100.0_{\pm 0.0}$ & $303_{\pm 49}$ & $0.55_{\pm 0.10}$ \\
\bottomrule
    \end{tabular}
}
    \end{threeparttable}
\end{table}
\section{Impact, Limitation and Future Work}
\label{limitation}

Deep learning holds tremendous potential across various domains. However, its applications must be secure and protect user privacy, which can be threatened by MIAs. Our research finds that when models have strong performance, existing defense methods are insufficient to resist the latest attack methods. Notably, our low-rank feature suppression strategy significantly increases the difficulty of MIAs. Therefore, we recommend integrating our defense strategy when fine-tuning pre-trained models for downstream tasks involving sensitive or private data.

Currently, our research, like most MIA-related studies, focuses on image classification tasks. In the future, it will be important to study other tasks and modalities, such as node prediction (vulnerable to graph reconstruction attacks \cite{zhou2023strengthening}) and autoregressive LLMs (susceptible to prompt inversion attacks \cite{zhang2024extracting}). Since our method is task-independent and can be applied to a variety of tasks and modalities, we expect that our defense method can provide stronger privacy protection in a wider range of application scenarios.

\end{document}